\documentclass[twocolumn, astrosymb]{aastex631}

\newcommand{\csb}{Pantheon+}
\newcommand{\csu}{CSP}

\usepackage{amsmath}

\graphicspath{{./}{figures/}}

\shorttitle{$H_0$ from blue SNe Ia}
\shortauthors{Gall et al.}

\begin{document}

\title{The Hubble Constant from Blue Type Ia Supernovae}

\correspondingauthor{Christa Gall}
\email{christa.gall@nbi.ku.dk}

\author[0000-0002-8526-3963]{Christa Gall}
\affiliation{DARK, Niels Bohr Institute, University of Copenhagen, Jagtvej 155A, DK-2200 Copenhagen N, Denmark}

\author[0000-0001-9695-8472]{Luca Izzo}
\affiliation{INAF, Osservatorio Astronomico di Capodimonte, Salita Moiariello 16, I-80131 Napoli, Italy}
\affiliation{DARK, Niels Bohr Institute, University of Copenhagen, Jagtvej 155A, DK-2200 Copenhagen N, Denmark}

\author[0000-0001-9666-3164]{Rados\l{}aw Wojtak}
\affiliation{DARK, Niels Bohr Institute, University of Copenhagen, Jagtvej 155A, DK-2200 Copenhagen N, Denmark}

\author[0000-0002-4571-2306]{Jens Hjorth}
\affiliation{DARK, Niels Bohr Institute, University of Copenhagen, Jagtvej 155A, DK-2200 Copenhagen N, Denmark}



\begin{abstract}
There is a persistent tension of about $5\sigma-6\sigma$ between the value of the Hubble constant, as derived from the local distance ladder vs. the cosmic microwave background, signaling either unaccounted for systematics in the measurements or `new physics', such as early dark energy. Determining the Hubble constant using Type Ia supernovae requires nontrivial and accurate corrections for dust extinction. 
To circumvent this obstacle, we here determine the Hubble constant from blue, and hence presumably unextinguished, supernovae. 
For two different 
sets of Type Ia supernova data and lightcurve fitting methods,
we find that when using blue supernovae only, the derived Hubble constant is consistently lower by 
$\sim$ 3 km s$^{-1}$ Mpc$^{-1}$ 
($70 \pm 2.1$ and $70.3 \pm 3.0$ km s$^{-1}$ Mpc$^{-1}$), 
and within 1$\sigma$ of the cosmic microwave background measurement, compared to when using all supernovae. 
This is consistent with the hypothesis that systematic effects in dust corrections may affect standard supernova cosmology. However, the number of blue calibrating Type Ia supernovae is small (about six), and values of the Hubble constant for a range of different supernova colors are consistent at the 1.2$\sigma$ level.
Upcoming major transient surveys will discover numerous unextinguished SNe~Ia, and thus be able to increase the precision of the Hubble constant measured from blue SNe~Ia, heralding a promising path toward resolving the Hubble constant tension. 
\end{abstract}



\section{Introduction} 
\label{sec:intro}

One of the most important scientific breakthroughs of the past few decades is the discovery of the acceleration in the expansion of the Universe  \citep{1998ApJ...507...46S,1998AJ....116.1009R,1999ApJ...517..565P}, i.e., the second derivative of the scale factor with respect to time. Despite this, nearly a century after the discovery of the expansion of the Universe \citep{1927ASSB...47...49L, 1929PNAS...15..168H}, there is disagreement about the true value of the present-day expansion rate, the Hubble constant ($H_0$, i.e., the first derivative of the scale factor). 
Tension at a significance of about  5$\sigma-6\sigma$ arises between different independent measurements  \citep{2019NatAs...3..891V, 2021CQGra..38o3001D, 2025JCAP...11..063C}. Most prominently, this is between that of `early Universe' measurements of $H_{0}$ = 67.36 $\pm$ 0.54 km s$^{-1}$ Mpc$^{-1}$ \citep{2020A&A...641A...6P} based on the cosmic microwave background (CMB) assuming a flat $\Lambda$ cold dark matter cosmological model and the local $H_{0}$ = 73.04 $\pm$ 1.04 km s$^{-1}$ Mpc$^{-1}$ measurements 
\citep{2016ApJ...826...56R, 2022ApJ...934L...7R}. 
The latter are based on the cosmic distance ladder involving observations of SNe~Ia and Cepheids, which are calibrated with geometric distances to the Large Magellanic Cloud \citep[][]{Pietrzynski2019}, Milky Way \citep{2021ApJ...908L...6R}, or the megamaser galaxy NGC 4258 \citep{2022ApJ...934L...7R, 2019ApJ...882...34F}.

In this paper, we address the measurement of $H_{0}$ using SNe Ia as standardizable candles. SN~Ia data sets cover a wide redshift range ($z < 2.26$) and are sufficiently large that statistical uncertainties are subdominant \citep{2014A&A...568A..22B, 2019ApJ...874..150B,2022ApJ...934L...7R, 2022ApJ...938..113S, 2022ApJ...938..110B, 2022MNRAS.514.5159M,2024ApJ...975...86V}. It is therefore important to investigate if unaccounted-for systematic uncertainties could be the source of the discrepancy. 
SN Ia cosmology requires a range of corrections related to intrinsic luminosity and color variations at optical and near infrared wavelengths 
\citep{1999AJ....118.1766P, 2009ApJ...704..629M, 2014ApJ...789...32B, 2018ApJ...869...56B, 2019PASP..131a4001P, 2023MNRAS.524..235D}, 
extinction corrections \citep{1996ApJ...473...88R, 2007ApJ...659..122J, Mandel17, 2011ApJ...731..120M, 2022MNRAS.510.3939M, 2021ApJ...909...26B, 2023MNRAS.524..235D}, 
correlations with host-galaxy properties \citep{2010ApJ...715..743K, 2010MNRAS.406..782S}, the progenitor systems and explosion mechanism of SNe Ia \citep{2018AA...611A..58G, 2018MNRAS.477..153A} or
astrophysical selection effects \citep{2021ApJ...912..150D, 2022Galax..10...24D, 2025JHEAp..4800405D}.

On the distance calibration, i.e., the anchoring of relative SN Ia distances, Cepheid variable stars \citep[][Leavitt Law]{1912HarCi.173....1L}, are often used to measure absolute distances to nearby SNe~Ia \citep{2001ApJ...553...47F, 2016ApJ...826...56R, 2018ApJ...869...56B, 2022ApJ...938...36R}. 
This method preferentially yields a high value of $H_0$, largely consistent with what is found using other methods, including Surface Brightness Fluctuations \citep[e.g.,][]{Tonry1998, 2021ApJ...911...65B, 2023ApJ...953...35G}, Tully--Fisher relation calibration \citep[e.g.,][]{1977A&A....54..661T, 2020AJ....160...71S}, or Fundamental Plane measurements \citep[e.g.,][]{2025ApJ...979L...9S}, which, however, typically report larger systematic uncertainties.
Measurements that are not in tension with the CMB value, e.g.,  $H_{0}$ of 69.85 $\pm$ 1.75 (stat) $\pm$ 1.54 (sys) km s$^{-1}$ Mpc$^{-1}$ 
is reached using the Tip of the Red Giant Branch method \citep{1995AJ....109.1645M, 2019ApJ...882...34F, 2021ApJ...919...16F, 2020ApJ...901..143U, 2025ApJ...985..203F}. 
Recently, with a new J-region asymptotic giant branch method, a value of $H_{0}$ = 67.96 $\pm$ 1.85 (stat) $\pm$ 1.90 (sys) km s$^{-1}$ Mpc$^{-1}$ has been found \citep{2024ApJ...961..132L}.

To determine the intrinsic brightness of an SN Ia, one needs to account for foreground dust extinction in its host-galaxy accurately. The degeneracy between the SN~Ia intrinsic colors and extrinsic reddening due to line of sight host-galaxy extinction \citep{Mandel17, 2019ApJ...874..150B, 2021ApJ...909...26B, 2022MNRAS.515.2790W} poses a significant challenge. Recent studies show that SNe~Ia with very red observed colors lead to a larger rms scatter between the data and the best-fit model
\citep[e.g.,][]{2021ApJ...909...26B, 2023ApJ...945...84P, 2021MNRAS.501.4861K,2022MNRAS.516.4822R,2023MNRAS.519.3046K}. 
Independent of the modeling and analysis approach, and whether or not additional correction steps (e.g., host-galaxy mass step) are included, the origin of this distribution is largely ascribed to variations in the total-to-selective-extinction parameter, $R_{V}$. 

The parameter $R_{V}$ essentially characterizes the dust properties along the line of sight to a supernova. 
In the Milky Way and neighboring galaxies \citep{1989ApJ...345..245C, Schlafly12, Schlafly16, 1999PASP..111...63F, Calzetti94, 2025Sci...387.1209Z} the value of $R_V$ ranges from around 2.5 to 6 for sight lines through the diffuse to dense interstellar medium (ISM), with a mean of 3.1 and a fairly small scatter for the Milky Way. 
In cosmological analyses of SNe Ia, one typically finds values ranging from 1.5 to 3, with a broad gaussian distribution \citep[e.g., mean $\sigma_{RV} > 1$,][]{2023ApJ...945...84P} and depending on the
priors adopted for SN~Ia intrinsic colors and reddening
\citep{2011ApJ...731..120M, 2014ApJ...789...32B, Mandel17, 2022MNRAS.510.3939M, 2022MNRAS.517.2360T, 2021MNRAS.508.4310T, 2023MNRAS.518.1985M, 2023MNRAS.519.3046K, 2024ApJ...975...86V, 2024arXiv240606215P, 2024MNRAS.531..953G}.
\citet{2021MNRAS.508.4310T} found values between 2.6 and 2.7, which are independent of the host-galaxy mass while other studies find a dependence on the mass of the host galaxies with larger mean $R_{V}$ values of $\sim$ 2.8 for lower mass galaxies and lower mean $R_{V}$ values of $\sim$ 1.5 for high mass galaxies
\citep{2021ApJ...909...26B, 2023ApJ...945...84P}.
From the lightcurves of individual SNe Ia, 
extremely low $R_{V} = 1.5$--1.8 values have also been inferred for a handful of highly reddened SNe~Ia, typically assuming a symmetric Gaussian prior for the SN~Ia intrinsic color 
\citep{Elias-Rosa06, 2008MNRAS.384..107E, 2008ApJ...675..626W, Amanullah14}.

A very small $R_{V}$ could point to special dust properties in the vicinity of SNe~Ia \citep{2008ApJ...686L.103G, 2018MNRAS.479.3663B}, although they can be difficult to reconcile with known physical dust properties \citep{2011piim.book.....D}. 
It remains puzzling that rarely larger $R_{V}$ values are inferred, contrary to those commonly found in the sight lines of the Milky Way, in an SN Ia calibrator galaxy \citep{2025arXiv250309702M}, or in samples of star-forming galaxies \citep{Calzetti94, Keel14, 2020ApJ...899..114D, 2021ApJ...922..135D}. 
Alternatively, the effect may be due to an inadequate treatment of SN~Ia colors \citep{2010A&A...523A...7G, 2011AJ....141...19B}. 
More fundamental issues inherent to differences in the intrinsic and extrinsic color distribution of different populations of SN~Ia in the different SN~Ia samples may also exist \citep[e.g.,][]{2025A&A...702A.176W, 2026arXiv260119854R}. 
As demonstrated by \citet{Mandel17, 2022MNRAS.510.3939M}, disentangling intrinsic color variations within the SN population from dust reddening to a cosmologically relevant precision remains challenging.  An entirely different approach to describe the intrinsic and extrinsic reddening effects may be needed \citep{2023MNRAS.525.5187W, 2024MNRAS.533.2319W, 2024MNRAS.534.2263P}. 

{In principle, SN cosmology would not depend on the details of dust corrections in the ideal situation where calibration and HF samples would be drawn from strictly identical SN and host-galaxy populations in an unbiased way. \citet{2022ApJ...934L...7R} carefully selected samples with similar properties (blue spiral galaxies). However, for precision cosmology, one cannot rule out second-order effects \citep{2024MNRAS.533.2319W}.

A promising avenue to eliminate the effects of reddening is to study SNe~Ia in the near-infrared (NIR), where dust extinction is suppressed relative to the blue--optical range. From modeling data in optical $+$ NIR, \citet{2023MNRAS.524..235D} found
$H_0$ = 74.82 $\pm$ 0.97 (stat) $\pm 0.84$ (sys) km ${\rm s}^{-1}\, {\rm Mpc}^{-1}$ when using the calibration with Cepheid distances to 37 host galaxies of 41 SNe~Ia, and 70.92 $\pm$ 1.14 (stat) $\pm 1.49$ (sys) km ${\rm s}^{-1}\, {\rm Mpc}^{-1}$ when using the calibration with tip of the red giant branch distances to 15 host galaxies of 18 SNe~Ia. \citet{2023A&A...679A..95G} found
$H_0$ = 72.3 $\pm$ 1.4 (stat) $\pm 1.4$ (sys) km ${\rm s}^{-1}\, {\rm Mpc}^{-1}$ in the J band and
$H_0$ = 72.3 $\pm$ 1.3 (stat) $\pm 1.4$ (sys) km ${\rm s}^{-1}\, {\rm Mpc}^{-1}$ in the $H$ band.

Considering the challenges in establishing a universally applicable and physical dust model for SN Ia cosmology, in this paper we take a complementary approach. We attempt to circumvent dust corrections entirely by only studying the bluest SNe Ia, which is presumably the least reddened subpopulation of SNe Ia \citep{Mandel17, 2023MNRAS.525.5187W, 2024MNRAS.533.2319W, 2023ApJ...945...84P, 2024arXiv240606215P}. We show that there is a dependence of $H_{0}$ on SN Ia color, with the bluest SNe Ia leading to values that are not strongly discrepant with the low CMB $H_{0}$ value (and at the same time not inconsistent with the SH0ES value). Section \ref{sec:datana} briefly describes the publicly available data sets and methods used in this work. In Section \ref{sec:bluetoredhubble} we demonstrate that lower values of $H_0$ are obtained for blue SNe in two different samples. We discuss the results and impact thereof on the $H_{0}$-tension in Section \ref{sec:discussion} and conclude that a promising way of dealing with uncertain dust corrections in addressing the $H_{0}$-tension may be to focus on unextinguished SNe Ia only.

\section{Data and Analysis} \label{sec:datana}

We use SN~Ia data that are acquired by different surveys, and for which SN~Ia lightcurve parameters are obtained from two distinct lightcurve fitters as described below. 
SN~Ia peak magnitudes are calibrated using common empirical relations between the SN~Ia peak luminosity and lightcurve parameters describing either the decline rate or shape of it \citep{1993ApJ...413L.105P} as well as the luminosity-color relation \citep{1998A&A...331..815T}. To obtain $H_0$, the intrinsic peak magnitudes and other model parameters (see Sect.~\ref{SS:comoset2}), we use Bayesian inference modeling to fit model parameters using Markov Chain Monte Carlo (MCMC) sampling.

\subsection{Pantheon+, SH0ES and {\tt SALT}}
\label{SS:comoset1}

We take SN~Ia lightcurve parameters from the recent comprehensive Pantheon+ SN Ia compilation together with the full cosmological model setup as described in \citet{2022ApJ...938..110B}. 
In alignment with this, we use the Cepheid distances to 42 calibrating SNe~Ia hosted in 37 individual galaxies as observed through the 
Supernovae and H0 for the Equation of State of dark energy
(SH0ES) program \citep{2016ApJ...826...56R, 2022ApJ...934L...7R}.

Pantheon+ comprises SN~Ia data taken from several SN~Ia surveys, such as the Harvard Smithsonian Center for Astrophysics \citep[CfA1--4][]{1999AJ....117..707R, 2006AJ....131..527J, 2009ApJ...700..331H, 2012ApJS..200...12H, 2009ApJ...700.1097H} and the Carnegie Supernova Project \citep[CSP-I][]{2010AJ....139..519C, 2010AJ....139..120F, 2011AJ....142..156S}, some are also compiled in the Open Supernova Catalog \citep{2017ApJ...835...64G}. Pantheon+ is the successor of the SuperCal supernova compilation \citep{2015ApJ...815..117S}, used by \citet{2016ApJ...826...56R} to obtain the Hubble constant measurement, which indicated for the first time a discrepancy ($3.4\sigma$) with the value derived from the Planck observations of the CMB.

The SN~Ia lightcurve parameters have been obtained through a 
retrained Spectral Adaptive Lightcurve Template (SALT2) 
model \citep{2007A&A...466...11G, 2010A&A...523A...7G, 2016ApJ...822L..35S, 2021MNRAS.504.4111T, 2022ApJ...938..111B}. 
The cosmological model is based on and adjusted to using SALT2 SN~Ia lightcurve parameters together with the \citet{1998A&A...331..815T} calibration. The distance modulus is computed as
\begin{equation}
    \mu = m_B - M_B + \alpha x_1 - \beta c - \delta _{\rm bias} + \delta _ {\rm host},
    \label{eq:eins}
\end{equation}
with $M_B$ the absolute $B$-band peak magnitude, and $x_1$ and $c$ the SALT2 specific lightcurve parameters characterizing the shape and color of the SN lightcurves, and $\alpha$ and $\beta$ as linear correction coefficients, which are fitted parameters in the cosmological model. 

Using SALT2 \citep{2005A&A...443..781G, 2007A&A...466...11G}, the color is defined as the rest-frame $B-V$ color with respect to the average at maximum $B$-band brightness as $c = (B-V)_{\mathrm{max}} - \langle B-V\rangle $. 
Then, $c$ represents the phase-independent total color difference of an SN (i.e., extrinsic extinction and intrinsic SN color variation) with respect to the mean color, assumed $\langle B - V\rangle =0$, of the SALT2 training sample. The latter is an inhomogeneous compilation of low and high redshift SNe~Ia from different
observations \citep[][and references therein]{2005A&A...443..781G, 2007A&A...466...11G}.

Besides the standard lightcurve shape and color correction parameters, $x_1$ and $\alpha$, additional corrections are included. The $\delta _{bias}$ accounts for observational selection biases determined from simulations and corrections determined from modeling of dust and supernova intrinsic color \citep{2021ApJ...913...49P, 2023ApJ...945...84P}. The term, $\delta _{\rm host}$, 
accounts for potential residual correlations between the host-galaxy mass
and the standardized brightness of SN~Ia
\citep[Eq. 4,][]{2022ApJ...938..110B}. The distance modulus uncertainty, 
for this model, accounts for uncertainties from observational selection effects, measurement uncertainties of lightcurve parameters, peculiar velocities, gravitational lensing, redshift measurements, and intrinsic variations of SNe~Ia. A detailed description of the uncertainties and associated covariances is presented in \citet[][and references therein]{2022ApJ...938..110B}.

For the purpose of this work, we adopt the already corrected SN~Ia peak magnitudes, $m_{B, \mathrm{corr}}$, as derived from 
the model of supernova standardization (Eq.~\ref{eq:eins}, \citet{2022ApJ...938..110B})
together with the covariance matrix for all SN~Ia corrected peak magnitudes and distance moduli, made publicly available\footnote{\label{note1}\url{https://github.com/PantheonPlusSH0ES/DataRelease/tree/main}}. 
The entire Pantheon+ data set consists of 1701 entries of in total, 1546 SNe~Ia, including the calibrator SNe~Ia within a redshift range 0.00122 $< z <$ 2.26137. 
To be consistent with $H_0$ measurements in the literature, we select the same Hubble flow (HF) SNe~Ia that obey the data selection criteria 
(SNe~Ia with late-type hosts similar to those of the Cepheid calibrators,
0.023 $ < z < $ 0.15; 
$|c| <$ 0.15; 
$|x_1| < 2$; 
$\sigma_{t_{PEAK}} <$ 5 days; 
$\sigma_{m_{B}} <$ 0.2 mag) 
as used by the SH0ES team and outlined in \citet{2022ApJ...934L...7R}.

We refer to this as the HF sample, which has a total of 270 entries for 234 individual SNe~Ia. 
For this work, we develop a custom-made {\tt Python} code  that makes use of publicly available {\tt Python} codes\footref{note1}
to select SNe~Ia and build respective covariance matrices for each color bin.

The corrected SN~Ia peak magnitude is defined as 
\begin{equation}
    m_{B, corr} = M_B + \mu, 
\end{equation}
where $\mu=\mu(z)$ is a cosmology-dependent distance modulus for SNe~Ia in the Hubble flow and $\mu=\mu_{\rm ceph}$ is a distance modulus measured from Cepheid observations for SNe~Ia in the calibration galaxies.

The combined likelihood as defined by \citet{2022ApJ...938..110B} is
\begin{equation}
    -2 \ln L = \Delta \vec{D}^T (C^{\rm cos}_{\rm stat+sys} + C^{\rm cal}_{\rm stat+sys})^{-1} \Delta \vec{D},
\end{equation}
with $D_i = \mu _i - \mu _{\rm ceph, i}$ for SNe, $i$, in the calibration galaxies, $D_i = \mu _i - \mu (z_i, H_0)$ for SNe, $i$, in the Hubble flow, where
\begin{equation}
\mu (z_{i}, H_0) = \mu _{\rm Planck}(z_{i}) + 5 {\rm{log}}_{10} \frac{H_0}{H_{0, \rm Planck}},
\label{eq:plank}
\end{equation}
leaving the absolute $B$-band magnitude, $M_B$, and $H_0$ as the only two free parameters to be fit. The statistical and systematic covariance for the cosmological sample is denoted $C^{\rm cos}_{\rm stat+sys}$ and $C^{\rm cal}_{\rm stat+sys}$ for the Cepheid calibrated SN~Ia host galaxies. 
The subscript Planck refers to quantities taken from the Planck cosmological model \citep{2020A&A...641A...6P} as implemented in {\tt python astropy}. To find the best-fit model parameters, we use the MCMC sampling technique as implemented in the {\tt Python} package {\tt emcee} \citep{2013PASP..125..306F}.

\subsection{CSP-I and CSP-II, SH0ES and {\tt SNooPy}}
\label{SS:comoset2}

We adopt the $B$-band SN~Ia lightcurve parameter of SNe~Ia lightcurve data compiled by CSP-I \& II \citep{2017AJ....154..211K, 2019PASP..131a4001P, 2020ApJ...901..143U} and the cosmological model from \citet{2024ApJ...970...72U}. 
All SNe~Ia lightcurve data have consistently been fit with the {\tt SNooPy} lightcurve fitter \citep{2011AJ....141...19B, 2014ApJ...789...32B} using the integrated `max-model' to obtain the peak $B$- and $V$-band magnitudes ($m_B$, $m_V$) and the lightcurve color-stretch parameter $s_{BV}$ as described also in \citet{2024ApJ...970...72U}. The SN~Ia colors are calculated as pseudocolors ($B_{max} - V_{max}$). 
We use the SN~Ia data, the calculated covariance matrix, and host-galaxy information as made available on GitHub\footnote{\url{https://github.com/syeduddin/h0csp}}. 
For consistency with the SN~Ia selection concept of \csb\ (Sect.~\ref{SS:comoset1}), 
we select HF SNe~Ia within a color range of the calibrator sample (i.e., $|(B-V)| < 0.22$ mag), a color stretch parameter $s_{BV} > 0.6$ and with a redshift $>$ 0.0233, which results in a final HF data set of CSP-I and CSP-II 177 SNe~Ia out of a total of 322 SNe~Ia used in \citet{2024ApJ...970...72U}. The $(B-V)$ and $s_{BV}$ parameter value ranges for the HF sample are chosen to be close to those encompassed by the SNe~Ia in the calibrator sample. The lower bound on $s_{BV}$ excludes subluminous, fast-declining SN~Ia subtypes \citep[e.g.,][]{2018AA...611A..58G}.
We find that out of the 177 HF SNe~Ia used here, only 29 SN~Ia are also in the \csb\ compilation (Sect.~\ref{SS:comoset1}) of which 27 are from the CSP-I sample. Due to this, the HF SN~Ia sample here can be considered an `almost independent data set'. 
We use the Cepheid distances from the SH0ES program \citep{2022ApJ...934L...7R} to 25 CSP SNe~Ia of \citet{2024ApJ...970...72U}. 

The apparent rest-frame $B$-band peak magnitudes are modeled as 
\begin{eqnarray}
 m_B = P^0 + P^1(s_{BV} - 1) + P^2(s_{BV} - 1)^2 + {}
 \nonumber\\
 {} R(B_{\rm max} - V_{\rm max}) + \zeta(M - \overline{M}) + \mu ,   
\label{eq:set2_mb}
\end{eqnarray}
where $P^0, P^1$ and $P^2$ are parameters of the polynomial $P^N(s_{BV})-1$ of the luminosity-decline rate relation up to second order ($N=2$). Additionally,
$R$ is the color correction parameter, which is the slope of the 
luminosity-color relation and $\zeta$ is the global slope parameter of the luminosity-host-galaxy stellar mass, $M = \log_{10}(M_\star/M_\sun)$, correction with $\overline{M}$, denoting the median value of the host-galaxy stellar mass of a sample of SNe~Ia. 
Then, $\mu = \mu _{\rm ceph}$ is the distance modulus of SNe~Ia in the Cepheid calibration sample. In order to obtain the Hubble constant, $H_0$, $\mu$ in Eq.~\ref{eq:set2_mb} is replaced by a distance modulus as a function of redshift and $H_0$ with
the model distance modulus for SNe~Ia is as defined in \citet{2018ApJ...869...56B},
\begin{eqnarray}
\mu (z_{\rm hel}, z_{\rm CMB}, H_0) = 5 {\rm{log}}_{10} \biggl[ \biggl( \frac{1+z_{\rm hel}}{1+z_{\rm CMB}} \biggr ) \frac{cz_{\rm CMB}}{H_0} {}
\nonumber\\
{} \biggl( 1 + \frac{1-q_{0}}{2} z_{\rm CMB} \biggr) \biggr ] + 25 ,
\end{eqnarray}
with the deceleration parameter, $q_0$=$-$0.53 and $z_{\rm CMB}$ and $z_{\rm hel}$ the redshifts relative to the CMB and the heliocentric frame of reference, respectively. 

As in Sect.~\ref{SS:comoset1}, we use $\mu = \mu _{\rm ceph}$ for SNe~Ia in the Cepheid calibration sample and $\mu = \mu (z_{\rm hel}, z_{\rm CMB}, H_0)$ to obtain $H_0$. 
The combined log-likelihood $\ln \mathcal{L} = \ln \mathcal{L} _{\rm cal} + \ln \mathcal{L} _{\rm cos}$ is then
\begin{equation}
        \Gamma_{X} = \frac{(m_B - m_{B, \rm model})^2}
        {\sigma_{tot}^2}, 
        \nonumber
\label{eq:set2_gam}
\end{equation} 
\begin{equation}
        \sigma_{tot} = \sigma_{X}^2 + \sigma_{\rm int}^2 + \sigma_{Y}^2, 
        \nonumber
\label{eq:sig}
\end{equation} 
\begin{equation}
    \ln \mathcal{L} _{X} = - \frac{1}{2} \sum_{i}^{N_{X}} \Gamma_{X,i} - \frac{1}{2} \sum_{i}^{N_{X}} \ln 2 \pi \sigma_{{\rm tot},i}^2 .
\label{eq:set1_Llc}    
\end{equation}
The subscript, $X$, refers to either the calibration or cosmology (i.e., HF SN~Ia) sample. 
Then, $m_{B}$ is the observed $B$-band magnitude for each supernova and $m_{B, \rm model}$ the intrinsic model magnitude as given in Eq.~\ref{eq:set2_mb}. 
The error term $\sigma_{X}^2$ represents 
the total uncertainties of all observed quantities and is as defined in Eq.~5 in \citet{2024ApJ...970...72U},
\begin{multline}
\sigma_{X}^2 = \sigma_{m_{B}}^2 + (P^1 + 2P^2(s_{BV} - 1))^2 \sigma_{s_{BV}}^{2}  \\
 + R^2 \sigma_{(B-V)}^2 - 2(P^1 + 2P^2 (s_{BV} - 1)) \, cov(m_{B}, s_{BV}) \\
+ 2R(P^1 + 2P^2 (s_{BV} - 1)) \, cov(s_{BV}, (B-V)) \\
 - 2R \, cov(m_{B}, (B-V)) + \zeta^2 \sigma_{M}^2. 
\label{eq:sigmaud}
\end{multline}
The error term $\sigma_{\rm int}^2$ is calculated for the SNe~Ia in the HF sample only and is the intrinsic scatter (included as a fit parameter). 
The error term $\sigma_{Y}^2$ $\equiv$ $\sigma_{\mu, \rm ceph}^2$ is used for the calibration sample and $\sigma_{Y}^2$ $\equiv$ $\sigma_{\rm pec}^2$ is the uncertainty due to peculiar velocities, $\sigma_{\rm pec} = 2.17 (v_{\rm pec} / cz_{\rm cmb})$, with $v_{\rm pec}$ representing the average peculiar velocity dispersion in the SN~Ia HF sample, as a free parameter.  
In total, there are eight model parameters to be fit ($P^0, P^1, P^2, R, \zeta, \sigma_{\rm int}, v_{\rm pec}$ and $H_0$). We assume Gaussian priors and we use the MCMC sampling technique as implemented in the {\tt Python} package {\tt emcee} \citep{2013PASP..125..306F}. 

\begin{deluxetable*}{lCCCccccccccccc}[htb!]
\label{tab:colordef}
\tablecaption{
 {\bf Sample Definition Criteria} }
\linespread{1.1}\selectfont\centering
\tablewidth{0pt}
\tablehead{
\colhead{Model\tablenotemark{a}} &
\multicolumn{3}{c}{SN~Ia Color Range\tablenotemark{b}} &
\multicolumn{3}{c}{\# SNe~Ia\tablenotemark{c}}   &
\multicolumn{4}{c}{$f_{SN}$\tablenotemark{d}} &
\multicolumn{3}{c}{\# SNe~Ia\tablenotemark{e}} \\
\colhead{} &   
\colhead{Blue} &
\colhead{Greeen}    &
\colhead{Red}     &
\colhead{Blue} &
\colhead{Greeen}    &
\colhead{Red}     &
\colhead{All}     &
\colhead{Blue} &
\colhead{Greeen}    &
\colhead{Red}     &
\colhead{\it Set 1} &
\colhead{\it Set 2} 
} 
\startdata 
\noalign{\smallskip}
\hline
\noalign{\smallskip}
Pantheon+  
& $C$ $<$ -0.1  & -0.1 $<$ $C$ $<$ 0.06 & $C$ $>$ 0.06 
& 6/28& 25/176 & 11/35  
& 17.9 & 21.4 & 14.5 & 32.4 
& 239 & 29 \\
CSP 
& $C$ $<$ -0.03  & -0.03 $<$ $C$ $<$ 0.13 & $C$ $>$ 0.13 
& 4/49 & 14/113 & 7/15  
& 14.1 & 8.2 & 12.4 & 46.7
& 29 & 177 \\
\enddata
\tablenotetext{a}{All SNe sets are selected in the redshift range 0.0233 $< z <$ 0.15 }
\tablenotetext{b}{For Pantheon+, $C$ is the SALT2 lightcurve fitting parameter $c$ while for CSP, $C$ denotes the ($B - V$) [mag] color.} 
\tablenotetext{c}{Number of SNe~Ia in the calibration / cosmology sample.}
\tablenotetext{d}{$f_{SN} = (N_{\rm calib} / N_{\rm cosmo}) \times 100 $, with $N_{\rm calib}$ the number of calibrating SNe~Ia and $N_{\rm cosmo}$, the number of SNe~Ia in the cosmological sample.}
\tablenotetext{e}{Total number of SNe~Ia that are in Pantheon+ ({\it Set 1}) and CSP ({\it Set 2}).}
\tablerefs{
\citet{2016ApJ...826...56R} 
\citet{2022ApJ...934L...7R} 
\citet{2022ApJ...938..110B}
\citet{2024ApJ...970...72U}}
\end{deluxetable*}

\begin{figure} 
\gridline{
          \hspace{-0.39cm}
          \fig{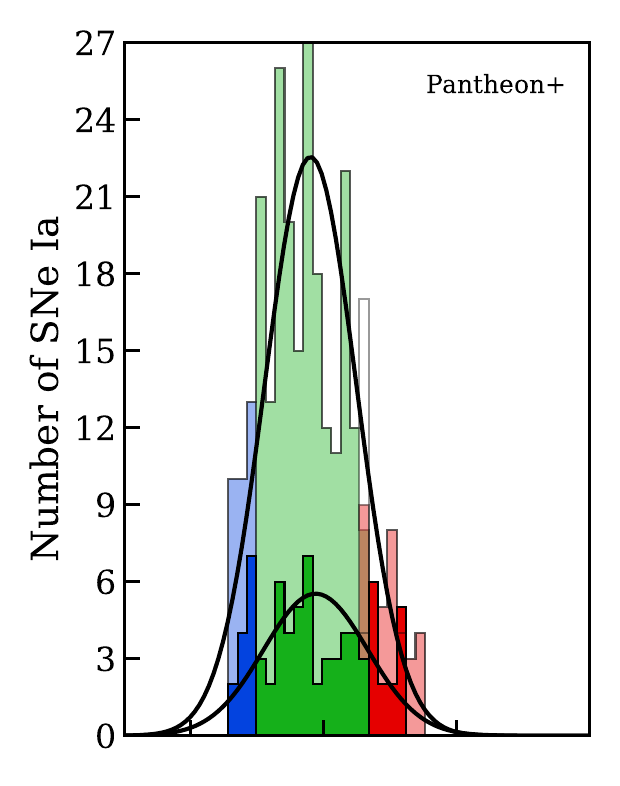}{0.25\textwidth}{}
          \hspace{-0.36cm}
          \fig{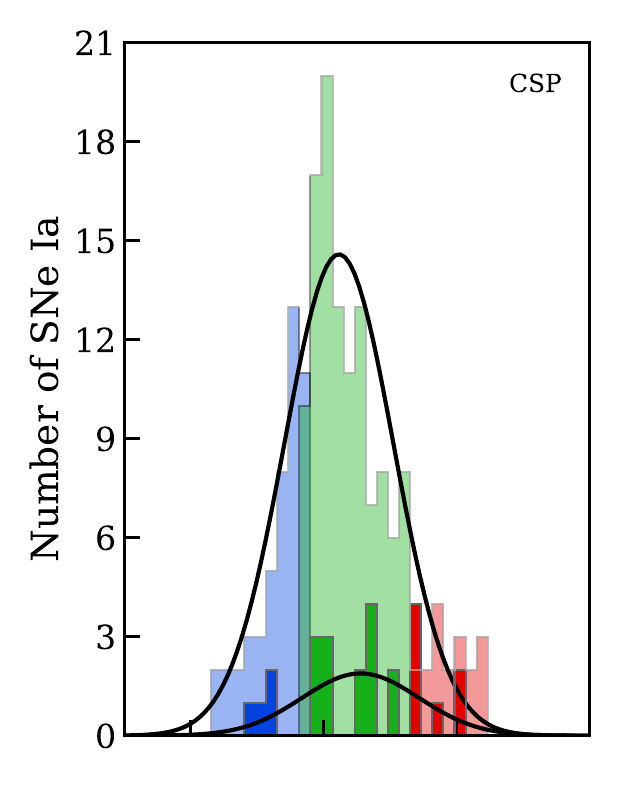}{0.25\textwidth}{}          }
          
\vspace{-1.2 cm}

\gridline{
          \hspace{-0.4cm}
          \fig{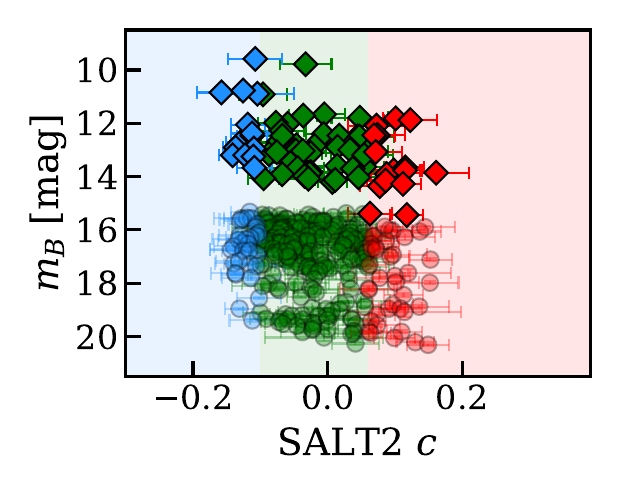}{0.25\textwidth}{(a)}
        \hspace{-0.35cm}
          \fig{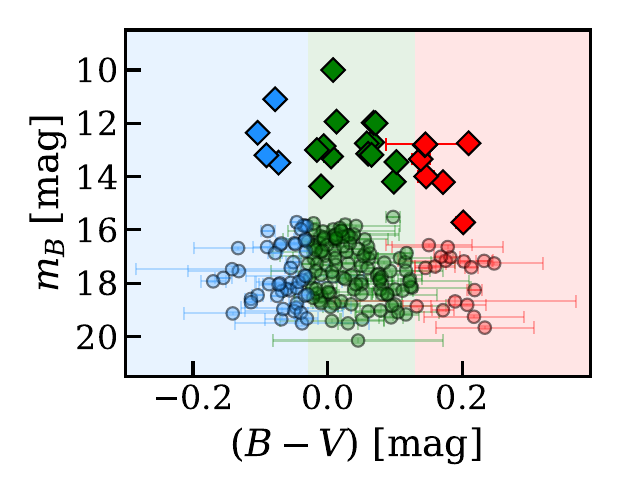}{0.25\textwidth}{(b)}
          }
\caption{Number of SNe~Ia (top row) and their measured peak magnitudes (bottom row) as a function of color. 
Shown are  SNe~Ia from Pantheon+ {\bf (a)} and SNe~Ia from CSP {\bf (b)}. Blue, green, and red colors mark the color range of the three color bins. The black solid lines are fit normalized probability density functions for each the HF and the calibrating SN~Ia sample. Solid-filled diamonds / histograms represent the calibration SNe~Ia while shaded filled circles / histograms represent SNe~Ia in the HF sample.   
}
\label{fig:distribution}
\end{figure}

\begin{figure*}[ht!]
\centering
\includegraphics[width=0.62\linewidth]{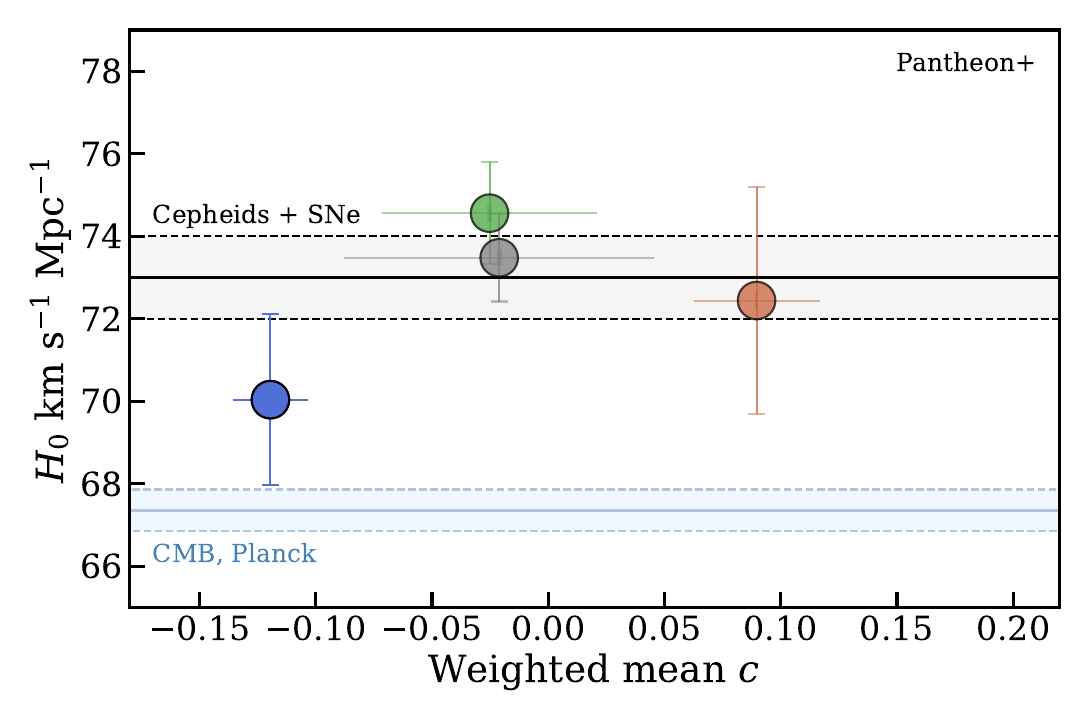}\\
\includegraphics[width=0.62\linewidth]{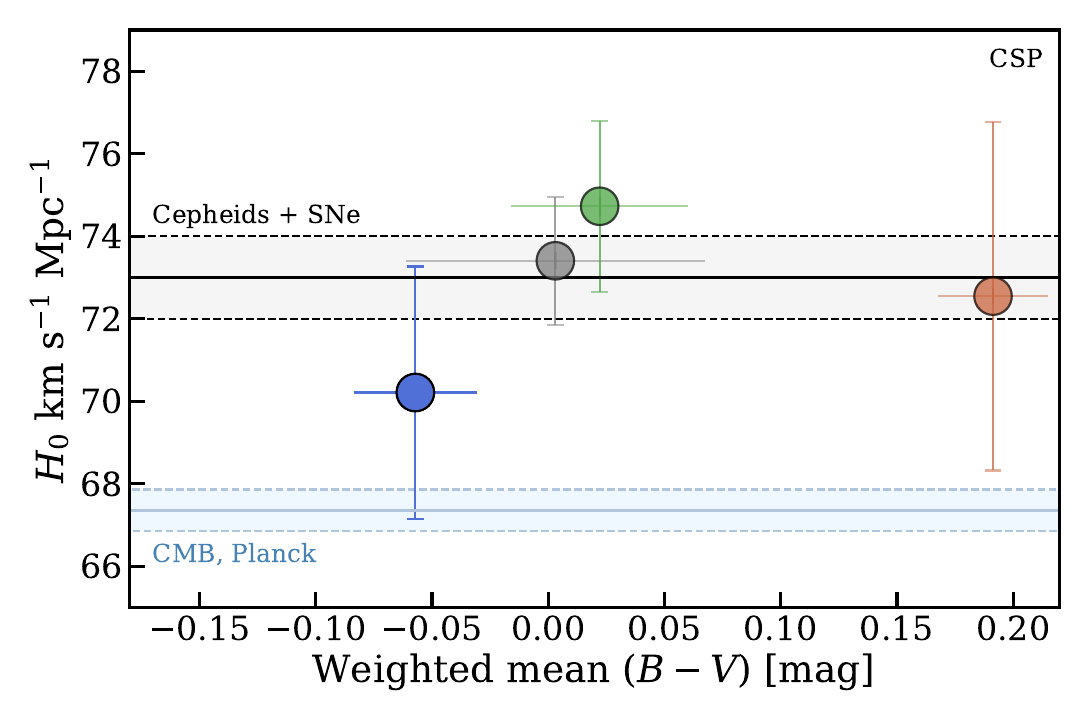}
\caption{$H_0$ as a function of color. {\it Upper panel:} shows $H_0$ for Pantheon+.  {\it Lower panel:} shows $H_0$ for CSP. Blue, green and red symbols mark the color range of the three color bins as defined in Table~\ref{tab:colordef}. The grey symbols refer to the measurement using all data. The weighted mean quantity (i.e., color) is shown with the weighted standard deviation to represent the scatter of the quantity instead of the error of the weighted mean, which is $< 0.1\%$. 
}
\label{fig:hnull1}
\end{figure*}

\section{The Hubble Constant from Blue Supernovae}
\label{sec:bluetoredhubble}

We define three discrete color bins into which we divide the calibration and HF SN~Ia data sets. 
For both \csb\ and \csu, the red and blue color cuts are chosen to contain only the bluest and reddest SNe~Ia in the tail of the color distribution of the entire calibrating SN~Ia sample. For \csb\ we consider all SNe~Ia with a measurement of $c < -0.1$ mag as blue, nearly unextinguished SNe~Ia. This is similar to the modeling-based estimate of the mean intrinsic color of the Pantheon+ sample \citep[$\overline{c} = -$0.077 $\pm$ 0.006;][]{2023ApJ...945...84P}.

However, Pantheon+ contains several duplicates, i.e, multiple entries of the same SN~Ia with lightcurve parameters derived from lightcurve data obtained by different telescopes and/or instruments. 
While any possible systematics arising from this are dealt with and accounted for in the statistical and systematic covariance matrix $C^{\rm cos}_{\rm stat+sys}$ and $C^{\rm cal}_{\rm stat+sys}$ \citep{2022ApJ...938..111B, 2022ApJ...938..113S}, some such SNe~Ia have color measurements with $c$ values over a wide range. 
For example, in \csb\ we find five SNe~Ia in each the calibration and the HF SN~Ia sample with $c$ measurements in both the blue and green color bin, and three SNe~Ia in the calibration and HF SN~Ia sample with $c$ measurements in both the green and the red color bin.
To avoid having the same SN~Ia in two color bins, we calculate the weighted mean color of SNe~Ia with multiple color measurements, and we use this weighted mean to decide in which color bin an SN~Ia can be placed. We then use the measurement of the respective SN~Ia in this color bin. The total number of SNe~Ia in each color bin is summarized in Table~\ref{tab:colordef}.

The CSP and SALT color definitions are not identical (see Sect.~\ref{sec:datana}). From the probability density function fits to the data compilations, as shown in Figure~\ref{fig:distribution} the distribution of the calibration sample peaks at $\bar{c}_{\rm cal} = -0.012 \pm 0.078$ mag for \csb\, while for \csu\ $\overline{B-V}_{\rm cal} = 0.057 \pm 0.14$ mag, which corresponds to a difference in mean peak color between the two distributions of about $0.07$ mag. To be in agreement with \csb, the blue color cut of \csb\ would translate into a blue color cut for \csu\ to $B-V = -0.03$ mag, and on the red side, $B-V = 0.13$ mag, which we adopt. Alternatively, the difference in the mean peak color between the two distributions of the HF samples is about $0.04$ mag. This translates into a \csu\ blue and red color bound of $B-V = -0.06$ and $B-V = 0.10$, respectively. Results are shown in Appendix~\ref{Apsec:panth}.

The HF sample is split into color bins with the same color cuts as the calibration sample. However, the color distributions of SNe~Ia in both the calibration and the HF samples are not identical. Thus, the ratio between the number of SNe~Ia in different color bins is different between the calibration and the HF samples. 
A summary is presented in Table~\ref{tab:colordef} and visualized in Figure~\ref{fig:distribution}. 

Figure~\ref{fig:hnull1} shows $H_0$ as a function of the color bin for our two samples. For $H_0$ there is a clear `blue-to-red' dependency on SN color in both cases. 
The upper panel shows that SNe~Ia (\csb) in the blue bin leads to a low 
$H_0 = 70.0 \pm 2.1$ km s$^{-1}$ Mpc$^{-1}$ 
while SNe~Ia in the green bin yield the highest $H_0$ values of 
74.5 $\pm 1.2$ km s$^{-1}$ Mpc$^{-1}$. 
SNe~Ia in the red bin also result in an
$H_0$ value of 72.5 $\pm$ 2.7 km s$^{-1}$ Mpc$^{-1}$. 
$H_0$ obtained with SNe~Ia in the green and red color bins are consistent with $H_0$ obtained for the entire sample, and both are consistent with standard SN~Ia $H_0$ values in the literature \citep{2016ApJ...826...56R, 2022ApJ...934L...7R}. All $H_0$ values can be considered consistent at a 1.2 $\sigma$ level. 

Nevertheless, the $H_0$ blue-to-red dependency on SN~Ia color is intriguing, considering that the corrected peak magnitudes and covariance matrix \citep{2022ApJ...938..110B} already include various corrections (see Sect.~\ref{SS:comoset1}). These are an SN~Ia color and dust-based color law which depends on host-galaxy properties \citep{2021ApJ...909...26B, 2021ApJ...913...49P, 2023ApJ...945...84P}.
Interestingly, using a different data compilation, lightcurve fitter, and SN~Ia color definition as described for \csu, a similar $H_0$ trend as shown in the lower panel of Figure~\ref{fig:hnull1} is obtained, with the lowest $H_0$ value of 70.3 $\pm$ 3.0 km s$^{-1}$ Mpc$^{-1}$ for blue SNe. 
All $H_0$ values for blue SNe~Ia in comparison to the $H_0$ values for the entire data sets are summarized in Table~\ref{tab:H0}.
We note that a small difference between the best-fit $H_0$ from the Pantheon$+$ likelihood and the SH0ES result reflects a difference in the corresponding likelihoods, which use distance moduli to the calibrators either as latent variables in a joint fit (SH0ES) or as fixed values in modeling the supernova data (Pantheon$+$).
In this paper we focus on blue SNe since we wish to explore the consequences of minimizing the effects of dust extinction. Nevertheless, we note that in both \csb\ and \csu\ samples, there is a downturn in $H_0$ for the reddest SNe (see Figure~\ref{fig:hnull1}). This downturn is less significant and more complex in that it may or may not be due to effects related to dust extinction.

%
\begin{deluxetable}{lcc}[htb!]
\label{tab:H0}
\tablecaption{
 {\bf $H_0$ from blue SNe} }
\linespread{1.1}\selectfont\centering
\tablewidth{0pt}
\tablehead{
\colhead{} &
\colhead{\csb}  &
\colhead{\csu}} 
\startdata 
\noalign{\smallskip}
\hline
\noalign{\smallskip}
Blue SNe~Ia~ 
& 70.0 (2.1) & 70.3 (3.0) \\ 
All data~ 
& 73.5 (1.1) & 73.4 (1.5) \\
\hline
\enddata
\tablecomments{Details of other $H_0$ values and model parameters are provided in Tables~\ref{tab:PANTHdata}~and~\ref{tab:UDDdata}. Units: km s$^{-1}$ Mpc$^{-1}$.}
\end{deluxetable}

One of the caveats of such a color bin test may be the so-called `Eddington bias' \citep{1913MNRAS..73..359E}. An SN~Ia with a color measurement falling on one side of the color cut with error bars extending into both color bins may, in reality, have a color that is either bluer or redder than the given color cut. This can introduce a bias in fitting the parameters, as also discussed in \citet{2008ApJ...686..749K} and \citet{2025ApJ...986..231R}. 
In Appendix~\ref{Apsec:csp}, we verify the effect by only considering SNe~Ia with measured colors and uncertainties, which are fully included in a bin. 
We find that only a very small percentage of SNe~Ia have color measurements with uncertainties extending into the neighboring color bin,
which does not significantly alter our results. We acknowledge that this does not exclude a possible bias by SNe~Ia that may 
have precisely but inaccurately measured lightcurve parameters. 
As discussed above, the choice of binning is aimed at identifying the bluest, least-extinguished SNe~Ia. Alternative choices of binning are explored in Appendix~\ref{Apsec:panth} and \ref{Apsec:csp} and shown in Appendix Figures \ref{fig:appcolbins}, \ref{fig:appNbins_fit} and \ref{fig:appcolbinsUDD}.

\section{Discussion} \label{sec:discussion}

The $H_0$ trend with SN color may signal a dependency on the prescription for disentangling the SN~Ia intrinsic color distribution and extrinsic color effects, and correcting for dust extinction. The differences in the response to dust models for the red SNe Ia (different approaches for \csb\ and \csu)
appear to strengthen this suspicion. 
Thus, taken at face value, the two compilations (\csb\ and \csu) of blue SNe Ia yield similar $H_0$ values of 
70.0 $\pm$ 2.1 km s$^{-1}$ Mpc$^{-1}$ and 70.3 $\pm$ 3.0 km s$^{-1}$ Mpc$^{-1}$, respectively,
which is consistent with their uncertainties with the CMB value (but also consistent with the standard analysis of these data sets). 
In principle, the selected subsamples of blue or red SNe~Ia may have different physical or environmental characteristics. We discuss a range of such possibilities below.

Alternatively, the $H_0$ dependency of SN color can be a statistical fluke and be dependent on the choice of binning due to the low number of SNe~Ia in each bin. However, as we demonstrate in Appendix~\ref{Apsec:panth}, the $H_0$ dependency on color is reproduced with different choices of binning, and the inferred $H_0$ values appear to increase systematically with SN color up to $c \sim$ 0 mag in both samples. The $H_0$ measurement obtained for the blue bin is a part of this systematic trend (see Sect.~\ref{ss:stats}, Appendix~\ref{Apsec:panth} and \ref{Apsec:csp}).

\begin{figure}[ht!]
\plotone{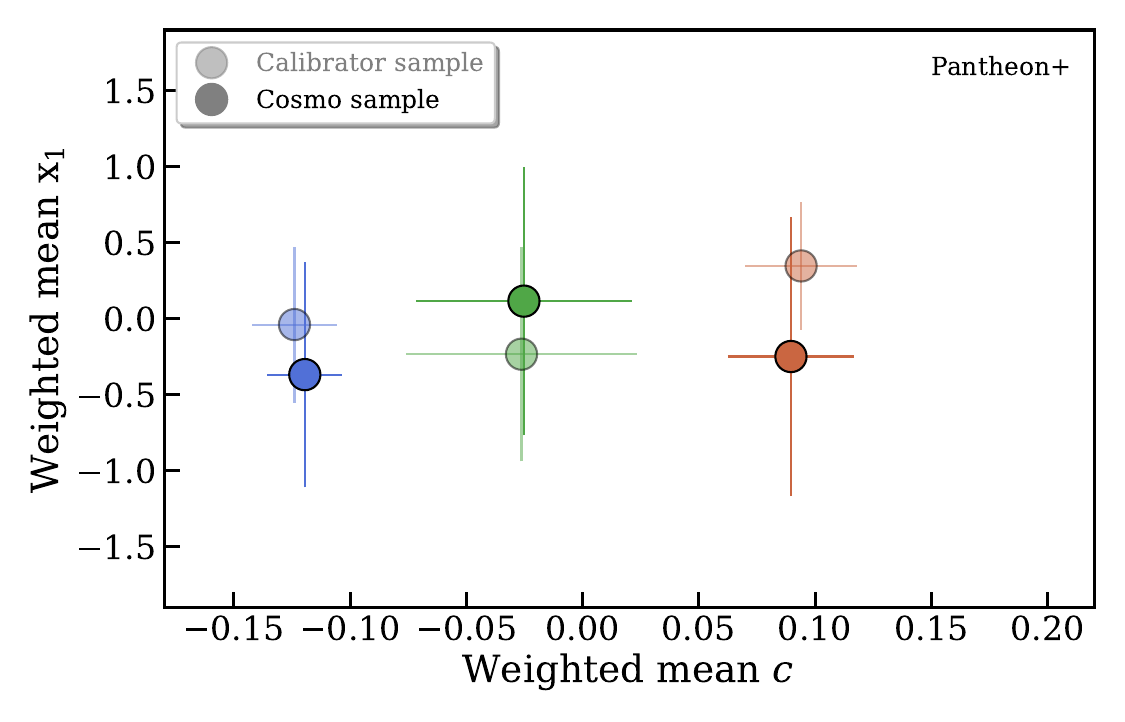}
\plotone{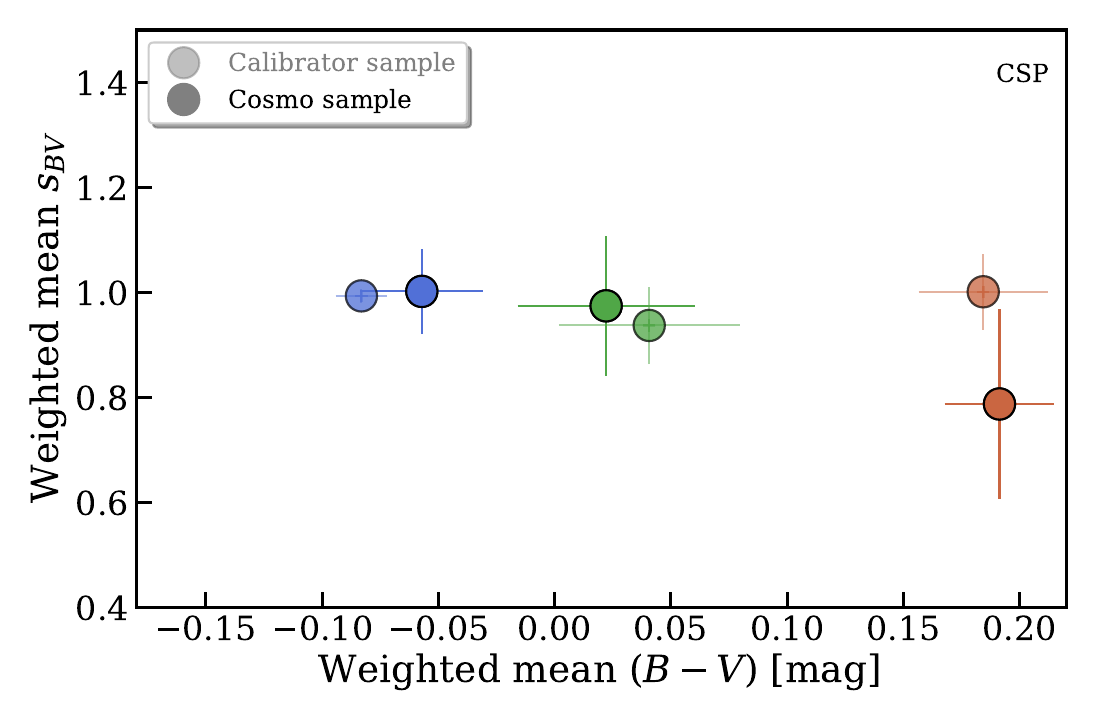}
\caption{Lightcurve shape parameter for different SN~Ia colors. {\it Upper panel:} shows $x_1$ for Pantheon+. {\it Lower panel:} shows $s_{BV}$ for CSP. Shown are the weighted standard deviations to represent the scatter of the quantities in each bin. The errors in the mean are smaller than the symbol sizes.}
\label{fig:stretch}
\end{figure}

\subsection{What are Blue SNe Ia?}
\label{ss:bluesne}

The empirical luminosity-color and luminosity-decline rate relations imply that intrinsically brighter SNe~Ia are bluer and have slower declining lightcurves \citep{1993ApJ...413L.105P, 1998A&A...331..815T}. Indeed, some of the brightest SNe~Ia, 91T-like SNe \citep{1992ApJ...384L..15F, 1992AJ....103.1632P} are found to have blue colors \citep{2018ApJ...864L..35S}. For the majority of `normal-bright' and `normal-declining' SNe~Ia, the colors at peak brightness are similar, with theoretical models able to produce SNe~Ia with blue colors \citep{2017ApJ...846...58H}. However, as long as the distribution of lightcurve shape parameters is the same in the different bins, this should not lead to differences in the inferred $H_0$.

Figure~\ref{fig:stretch} shows that the SNe~Ia in the blue bin in the \csb\ sample have a weighted mean SALT2 lightcurve shape parameter consistent within their uncertainties with that of SNe~Ia in the green and red bins. This is valid for both the calibration and the HF SN~Ia sample. The \csu\ SNe~Ia in the red bin appear to have slightly lower $s_{BV}$. 
There is no obvious trend, but if anything, they are opposite in the two cases. Thus, there is no indication of clear differences among the three bins, in either sample, and we conclude that the lower $H_0$ for blue SNe Ia is unlikely due to lightcurve differences among the bins.

\begin{figure}
\plotone{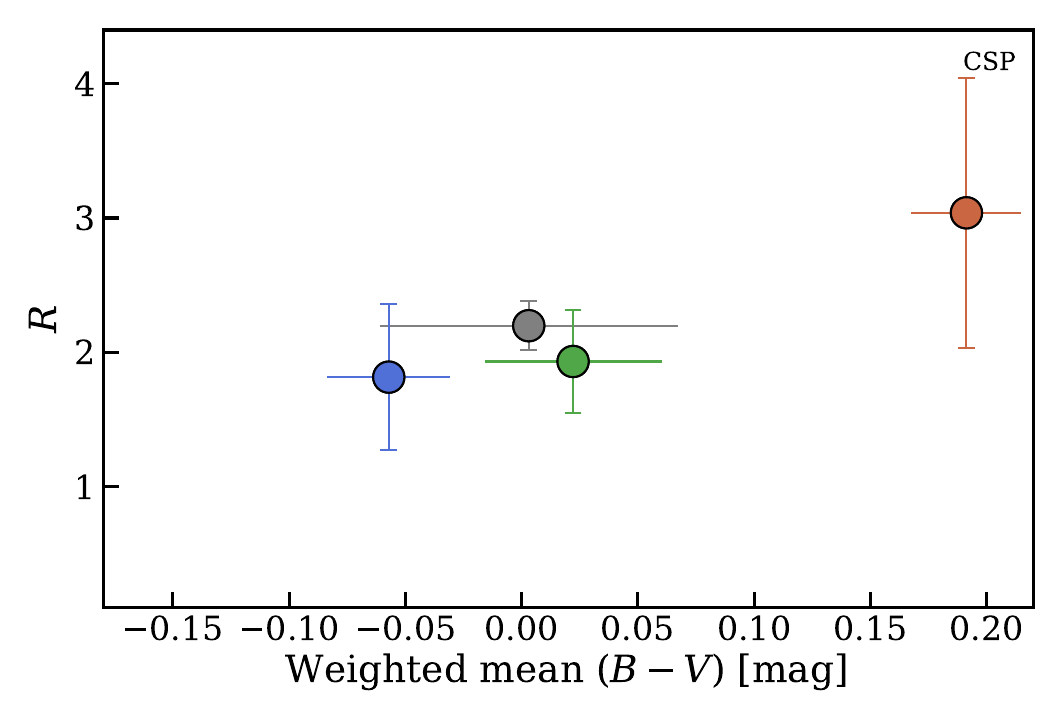}%
\caption{Color correction parameter for different SN~Ia colors. 
Here, $R$ for the CSP sample obtained for the blue, green, and red bins, including $R$ when using all data (gray symbol). The weighted standard deviations are shown to represent the scatter of color in each bin. The errors in the mean are smaller than the symbol sizes.
}
\label{fig:beta}
\end{figure}

For unextinguished SNe~Ia, the color correction parameters, $\beta$ or $R$, reduce to a correction coefficient of the intrinsic luminosity-color relation. Thus, 
a dependence of the local slope on
SN~Ia color may be expected if dust extinction has not been correctly accounted for \citep{Mandel17}. 

Indeed, smaller values of $\beta$ have been obtained for bluer SNe~Ia than for redder SNe~Ia \citep{2010ApJ...716..712A, 2021MNRAS.508.4656G}, with values of $\beta$ typically $\lesssim 2.6$  \citep[although see][for very red supernovae]{2022MNRAS.516.4822R}.
Such small $\beta$ values are unlikely due to dust alone ($R_V \approx \beta - 1.0$), as such steep extinction curves $(R_V \lesssim 2$) are inconsistent with observed dust properties in the ISM of galaxies (see Sect.~\ref{sec:intro}).
For \csb, $\beta$ as a function of SN~color is accounted for by the employed color model, which is trained on SNe~Ia at $z > 0.03$ and adopted for the calibration sample. 
However, for \csu, we can test if $R$, which can be interpreted similarly to $\beta$, changes with SN color. Figure~\ref{fig:beta} shows $R$ as a function of color.
Indeed, we see the expected trend of $R$ values increasing with color as SNe Ia become redder and hence more reddened.

Figures~\ref{fig:stretch} and \ref{fig:beta} are therefore consistent with bluer supernovae being less reddened. 
We conclude that SNe~Ia  with $c < -0.1$ (blue bin) in \csb\ likely are dominated by SNe~Ia close to their intrinsic peak magnitudes and colors. Consequently, SNe~Ia with measured $c > -0.1$, must be more strongly affected by dust \citep{2023MNRAS.525.5187W, 2024MNRAS.533.2319W, 2024MNRAS.534.2263P, 2024ApJ...975...86V}. 

\begin{figure}[ht!]
\plotone{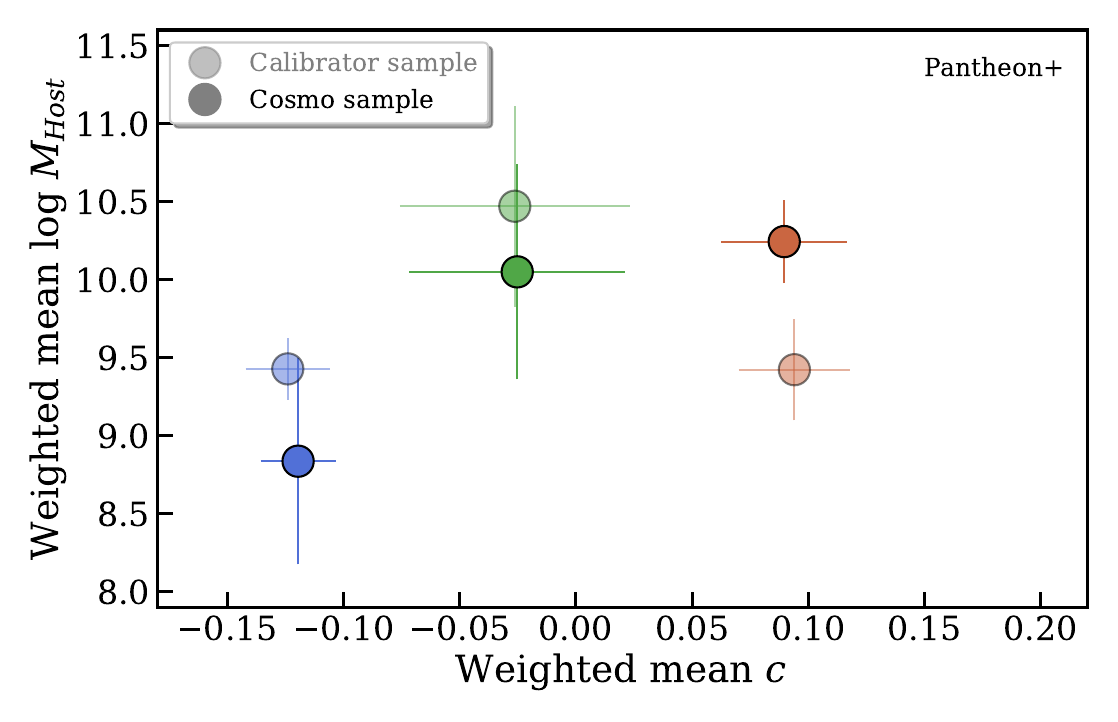}
\plotone{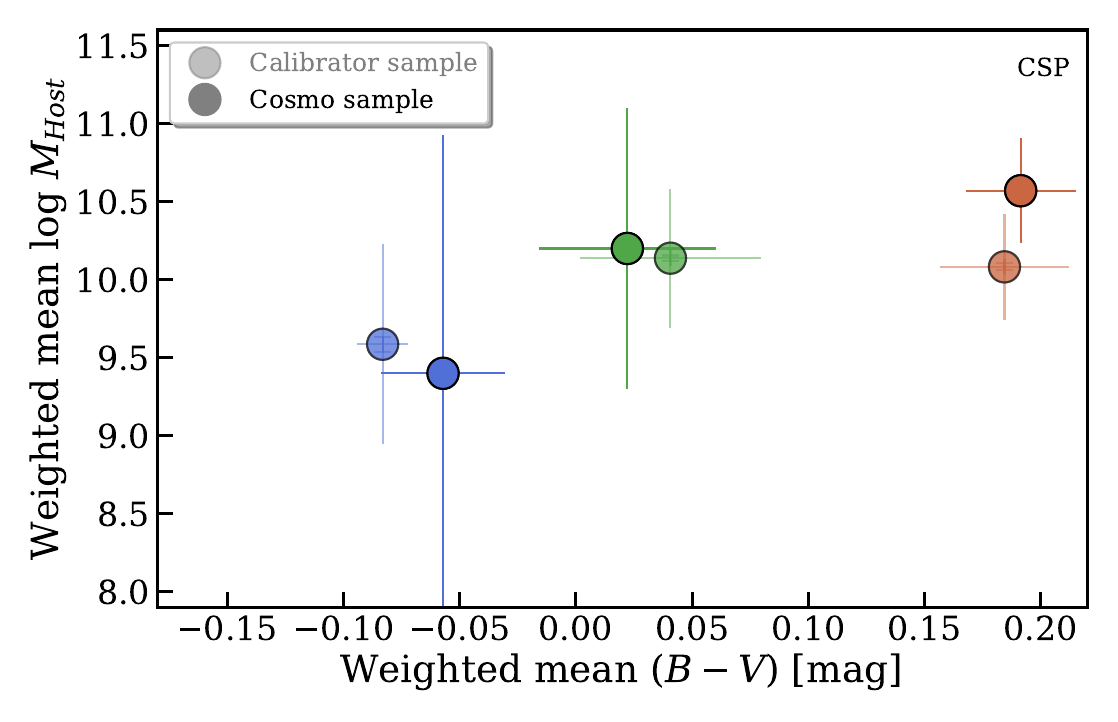}
\caption{Host-galaxy mass distribution with color. {\it Upper panel:} shows log $M_{\rm host}$ for the Pantheon+ data compilation. Host-galaxy masses are taken from the online tables of both data compilations (Sect.~\ref{sec:datana}). We note that host-galaxy masses are not available for all hosts of the \csb\ data compilation. The fractions of host galaxies with measurements for SNe~Ia in the calibration sample are 0.3 (blue), 0.64 (green), and 0.64 (red), and for SNe~Ia in the cosmological sample are 0.8 (blue), 0.93 (green), and 0.97 (red). Shown are the weighted standard deviations to represent the scatter of the quantities in each bin. The errors in the mean are smaller than the symbol sizes.
{\it Lower panel:} shows log $M_{\rm host}$ for CSP.}
\label{fig:mass}
\end{figure}

\begin{figure}[ht!]
\epsscale{1.2}
\plotone{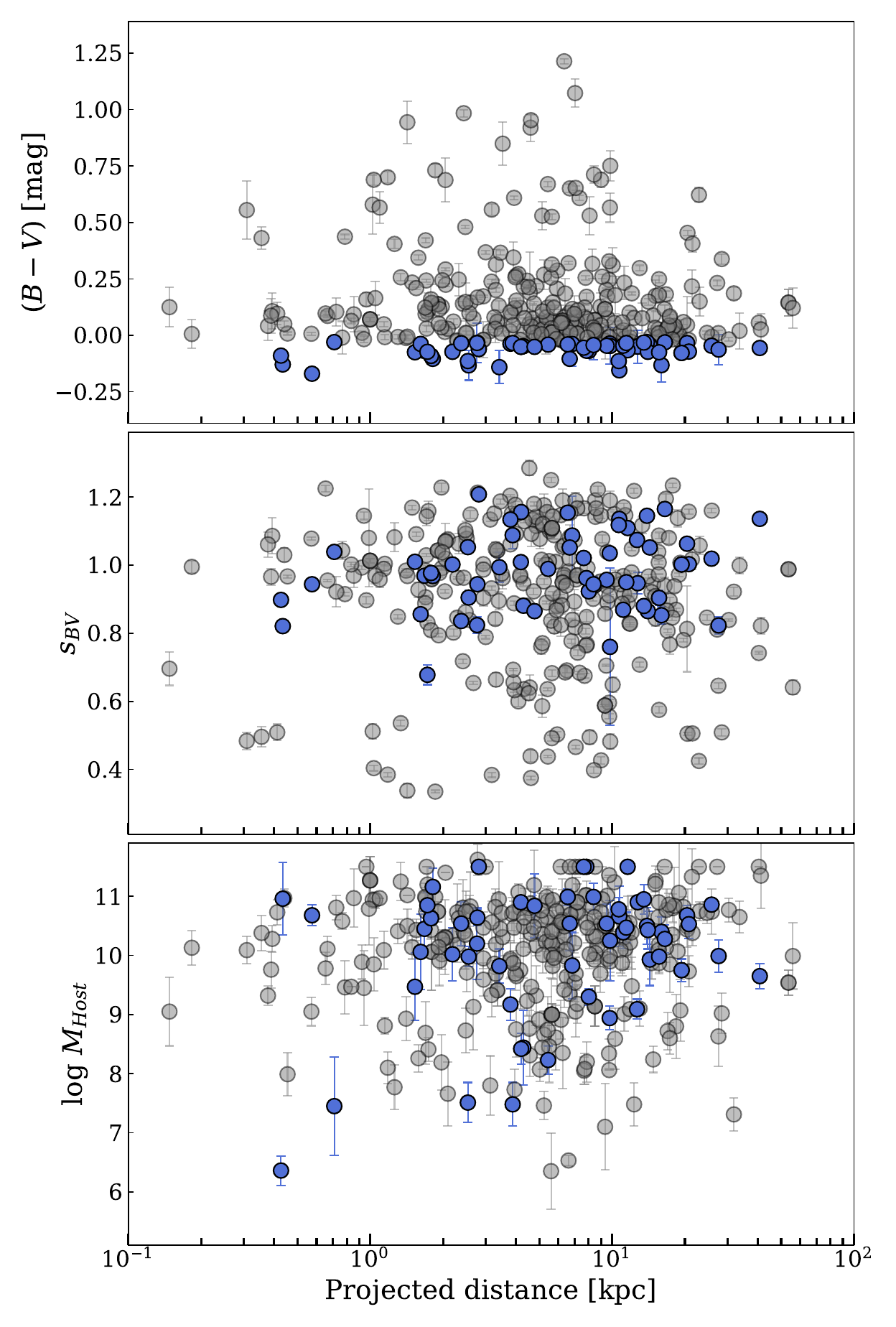}
\caption{Projected distance of SNe~Ia in CSP. Highlighted in blue are  SNe~Ia in the blue color bin ($B-V < -0.03$ mag). {\it Upper panel:} SN~Ia color. {\it Middle panel:} light curve shape parameter. {\it Lower panel:} host-galaxy mass. }
\label{fig:projection}
\end{figure}

\subsection{Blue SNe Ia and their Locations in Galaxies}

The majority of normal-bright and normal-declining SNe Ia with both blue and red colors occur in late-type galaxies. This is the case for a subsample of the Pantheon+ SNe~Ia \citep{2020MNRAS.499.5121P}, and all SNe~Ia with Cepheid distances  \citep{2016ApJ...826...56R, 2022ApJ...934L...7R}. Typically, late-type galaxies are gas- and dust-rich spirals with ongoing star formation \citep{1983ApJ...272...54K}. 

Early-type galaxies also host normal SNe~Ia, including the so-called `transitional' SNe~Ia. These events decline faster than the standard normal SNe~Ia but are not as faint as SN~1991bg-like SNe~Ia \citep{2008MNRAS.385...75T, 2018AA...611A..58G, 2023MNRAS.522.4444H}. These SNe~Ia have $B-V$ color curves that are those of normal SNe~Ia alike, but some such transitional SNe~Ia can have blue colors at peak brightness \citep[e.g., ][]{2013MNRAS.430..869S, 2018AA...611A..58G, 2026A&A...706A.381I}.  Interestingly, galaxies with SBF distances are primarily early-type galaxies \citep{2021AA...647A..72K, 2021ApJS..255...21J}. 

Studies of SN~Ia samples show that SNe~Ia in both early- and late-type galaxies have, on average, a similar SN~Ia color distribution (e.g., mean  SALT $c \sim 0$), although SNe~Ia in early-type galaxies appear to be marginally bluer \citep{2010MNRAS.406..782S, 2017NewA...51...43H, 2020MNRAS.499.5121P}. After lightcurve shape and color corrections, SNe~Ia in early-type galaxies, or galaxies with low specific star formation rates, high stellar masses, and high metallicity, are found to be brighter than those in late-type galaxies \citep{2008ApJ...685..752G, 2009ApJ...700.1097H, 2010MNRAS.406..782S, 2010ApJ...715..743K, 2021AA...647A..72K}. Other recent works show that by splitting the SN~Ia sample of the 5-year Dark Energy Survey into two color bins with $c < 0$ and $c > 0$, a lower r.m.s. scatter in the Hubble residuals can be achieved for SNe~Ia in the bluer bin, which are also hosted in less massive and blue $U-R$ colored galaxies \citep{2023MNRAS.519.3046K}.  

Figure~\ref{fig:mass} shows that \csb\ 
SNe~Ia in the blue bin appear to have less massive hosts than SNe~Ia in the redder bin, although there are only a few SNe~Ia with host-galaxy mass measurements.
For massive galaxies, there is a positive correlation between extinction and star-formation rate. However, for less massive galaxies ($< 10^{10}$ solar masses), \citet{2013ApJ...763...92Z} found that extinction is anticorrelated with star-formation rate. Thus, one can achieve low extinction for low-mass, star-forming galaxies, consistent with blue SNe being located in low-extinction environments.

Moreover, we calculate the projected distances of the CSP SNe~Ia as in \citet{2024ApJ...970...72U}. As shown in Figure~\ref{fig:projection}, there is no obvious preference for the projected distance of blue SNe~Ia relative to redder SNe Ia, although this does not imply that blue SNe~Ia are at the same location as redder SNe~Ia, e.g., close to the center of galaxies.
Indeed, \citet{2024AJ....167..131P} show that SNe~Ia with SALT2 $c < 0$ mag are missing in the inner regions of disk-dominated galaxies, which are home to SNe~Ia with $c > 0$ mag. While a link between the SN~Ia age and the age difference between the disk and bulge is ruled out as the origin of the missing blue SNe~Ia, significant dust extinction in the inner region of the disk is likely to redden the SNe~Ia. 
Since SNe with SALT2 $c < -0.1$ mag are unlikely to be strongly affected by extinction, the sight lines towards blue SNe must largely be dust free, independent of their projected distance. 
We suggest that blue, nearly unextinguished SNe~Ia are likely located in the outskirts and halos of galaxies where dust column densities can be low.

\subsection{Complexity of Dust Extinction}

Figure~\ref{fig:hnull1} and Appendix Figures~\ref{fig:appcolbins}, \ref{fig:appNbins_fit} and \ref{fig:appcolbinsUDDM24} show that independent of data sets, dust models, and lightcurve fitters, `red' SNe ($c > -0.1$, $(B-V) > 0.015$ mag) lead to high $H_0$ values consistent with \citet{2022ApJ...934L...7R} \citet{2026ApJ...998..101P} within uncertainties.
Since red SNe~Ia are either red and faint intrinsically or they are brighter and bluer SNe~Ia that are affected by dust extinction, unaccounted-for systematics of both color effects are plausible reasons for the $H_0$ variations.    
As shown by \citet{2022MNRAS.515.2790W}, simply accounting for different $\beta$ for the calibration and the cosmological sample leads to smaller values of $H_0$ than favored in the literature using a global $\beta$. \citet{2022MNRAS.515.2790W} found that $\beta$ appears to be larger for SNe in the calibration sample than the SNe in the cosmological sample. 
This points to either differences in dust extinction of SNe~Ia in the two samples, or differences in the intrinsic color distribution of SN~Ia in each of the samples. 
While this could be the result of an incorrect 
dust correction method
\citep{2015ApJ...815..117S}, the $H_0$ variations with SN~Ia color noticed in both Pantheon+ and CSP (see Figure~\ref{fig:hnull1}) suggest that the full complexity of dust extinction of SNe~Ia may not be captured by any of the dust models employed in lightcurve fitters and cosmological models.

Indeed, recent 3D Milky Way $R_V$ mapping reveals complex patterns throughout the different ISM regions, with large ($>$ 3.1) $R_V$ values found in both high and low density ISM regions, surprisingly low $R_V$ values in moderate density regions, and a strong correlation between high $R_V$ values and star formation \citep{2025Sci...387.1209Z}.  
Measurements of reddening curves through single-cloud and translucent cloud sight lines of the diffuse ISM of the Milky Way result in $R_V = 3.1 \pm 0.4$ \citep{2019ApJ...886..108F, 2023A&A...676A.132S}, in agreement with the mean $R_V$ value of the Milky Way. On the other hand, $R_V > 3.1$ are typically found through sight lines of the dense and cold ISM \citep[e.g.,][]{2019ApJ...886..108F}. 
In some such single-cloud sight lines, and strongly star forming regions in the Large Magellanic Cloud bar, a large `dark dust' component of up to 1--3 mag has been found \citep{2020ApJ...899..114D, 2020A&A...641A..35S, 2021ApJ...922..135D}.
Dark dust, also known as gray dust, is wavelength-independent dust originating from very large grains. Such dust leads to flat extinction curves that do not affect the colors of an extinguished object but affect the brightness of the object, and can add further complexity to dust extinction corrections of SN~Ia.

While the majority of host-galaxy sight lines toward SNe~Ia are likely a combination of different ISM regions, possibly resulting in average $R_V$ values \citep{2020A&A...641A..35S, 2023A&A...676A.132S},
some sight lines may be dominated by a certain ISM region, ISM density, or dust composition with very individual $R_V$ values. 
Given this inhomogeneous and complex nature of dust along the line of sight to SN~Ia and that SNe~Ia do not have a preferred location in galaxies, employing global host-galaxy (e.g., extinction, mass) corrections is likely to lead to under- or over-estimations of extinction, and hence SN~Ia magnitudes and colors.

\subsection{Low Number Statistics}
\label{ss:stats}
The $H_0$ values measured from SNe~Ia in the blue bin can be argued to be a statistical fluctuation. This is, because the number of calibrating and HF SNe~Ia (see Table~\ref{tab:colordef}) in the blue bin is low.  
However, the calibrating-to-HF SNe~Ia ratio in the blue bin is comparable to, if not larger than, 
the calibrating-to-HF SNe~Ia ratio of the entire sample (e.g., Pantheon+ blue bin). The same is the case for the green and red bins (see Table~\ref{tab:colordef}).
Hence, we tested this possibility in Appendix~\ref{Apsec:panth}, where we split the Pantheon+ sample into six color bins with a constant $\Delta c = 0.053$ mag. Each color bin has a similar number of SNe~Ia as the blue bin. It is expected that $H_0$ values derived from these bins would fluctuate randomly around a mean value. Figure~\ref{fig:appcolbins} shows that this is not the case. Rather, there is a gradual increase of $H_0$ with color, reaching a high $H_0$ value $> 76$ km s$^{-1}$ Mpc$^{-1}$ around $\bar{c} \approx 0$ mag. As also shown in \citet[][Figure~1]{2024MNRAS.533.2319W} and noticed in Figure~\ref{fig:mass}, such high $H_0$ values are driven by SNe~Ia in high-stellar-mass host galaxies ($M_{\star} > 10^{10}$ M$_{\sun}$), for which dust extinction is likely underestimated for reddened SNe~Ia.
We also tested this for the CSP sample, which is split into four color bins with $\Delta (B-V) = 0.08$ mag (Appendix~\ref{Apsec:csp}) and find a similar trend of $H_0$ with color, i.e., an increase of $H_0$ from blue toward redder SNe~Ia and a decrease of $H_0$ at the reddest end, rather than random fluctuations. 

We note that while statistical fluctuation causing the low $H_0$ may be unlikely, $H_0$ is nevertheless sensitive to the low number of SNe~Ia, their lightcurve parameter estimations, and extinction corrections. The latter may be affected by possibly only a few, unextinguished SNe~Ia in the lightcurve fitter (e.g., SALT2) training samples. 
With upcoming major transient surveys on the horizon, such as the Vera C. Rubin Observatory Legacy Survey of Space and Time and other simultaneous, complementary photometric and spectroscopic surveys, the number of blue, nearly unextinguished SNe~Ia will increase. This will allow a robust measurement of $H_0$ derived entirely from blue SNe~Ia and thus, independent of dust corrections.

\section{Conclusion}

In this paper, we used different SN~Ia data sets for which SN~Ia lightcurve parameters, such as lightcurve shape and color parameters, have been obtained with different lightcurve fitting methods (SALT2 and {\tt SNooPy}). 
We used data compilations from Pantheon+ and CSP \citep{2022ApJ...938..110B, 2022ApJ...938..111B, 2022ApJ...938..113S, 2024ApJ...970...72U} together with Cepheid distances  \citep{2022ApJ...938...36R}.

The key findings are summarized below:
\begin{itemize}
    \item Independent of the data set, lightcurve fitter, and cosmological model investigated here, we find low $H_0$ values ($\sim$ 70 km s$^{-1}$ Mpc$^{-1}$) for nearly unextinguished blue SNe~Ia. 
    \item SNe~Ia with colors $c > -0.1$ mag tend to result in higher $H_0$ values ($H_0 > 72$ km s$^{-1}$ Mpc$^{-1}$). 
    \item The $H_0$ values from nearly unextinguished blue SNe~Ia are within 1.2 $\sigma$ of $H_0$ values from extinguished SNe~Ia (green bin) and within 1 $\sigma$ of those measured from the CMB. This is a net effect of (i) a smaller best-fit value ($\sim$3 km s$^{-1}$ Mpc$^{-1}$) and (ii) larger errors due to smaller sample size. We stress that our results (see Figure~\ref{fig:hnull1}) are fully consistent with previously published results of high $H_0$ values from the same data \citep{2022ApJ...934L...7R, 2024ApJ...970...72U}. The errors on our $H_0$ values from the blue (and red) SNe~Ia encompass this 'high' value of $H_0$. At the same time, the lower mean $H_0$ value and the larger error associated with it are also consistent with a 'low' $H_0$ value.

    \item There seems to be no obvious difference of the projected distance of blue SNe~Ia compared to redder SNe~Ia although blue SNe~Ia must be located in low-extinction environments. 
\end{itemize}
 
Our findings suggest that the majority of SNe~Ia in cosmological samples may be affected by more complex dust extinction than what is captured by dust models currently employed in the standardization of SNe~Ia and cosmological models. 
As directly measuring extinction for each SNe~Ia in distant galaxies is challenging, turning to only blue, presumably unextinguished, SNe~Ia may be a promising path for future cosmology.  Including reddened supernovae requires careful modeling of SNe~Ia populations in relation to their environments. Along these lines, recent improvements in matching SNe~Ia in the Hubble flow against the calibration SNe~Ia resulted in as low values of the Hubble constant as the estimate obtained in this study from blue SNe~Ia \citep{Wojtak2025, 2025arXiv251114332M}.

\begin{acknowledgments}
This work is supported by a VILLUM FONDEN Young Investigator Grant (project No. 25501), a Villum Experiment grant (VIL69896), and by research grants (VIL16599, VIL54489) from VILLUM FONDEN. 
We thank the anonymous referee for constructive comments, and we thank 
Stephen Thorp, Gautham Narayan, 
and Chris Burns for helpful discussions. 
\end{acknowledgments}

\software{{\tt astropy} \citep{2013A&A...558A..33A, The_Astropy_Collaboration_2022}, {\tt emcee} \citep{2013PASP..125..306F}
          }
\bibliography{OneA}{}
\bibliographystyle{aasjournal}

\appendix

\section{The Pantheon+ sample, testing color dependence}
\label{Apsec:panth}

Here we show the result for different selections of color bins using \csb. First, we split the sample into two bins, $c < 0$ and $c > 0$.  
Second, we calculate a $\Delta c = (c_{\rm cal, max} - c_{\rm cal, min}) / N_{\rm bin}$ of the entire color range covered by the calibration sample for different numbers of bins, $N_{\rm bin}$ = 2, 3, 4, 5, and 6, with $\Delta c = 0.133, 0.088, 0.066, 0.053$ and $0.044$, respectively.   
The results for both experiments are shown in Figures~\ref{fig:appcolbins} and \ref{fig:appNbins_fit}.

\begin{figure}[ht!]
\epsscale{1.0}
\plottwo{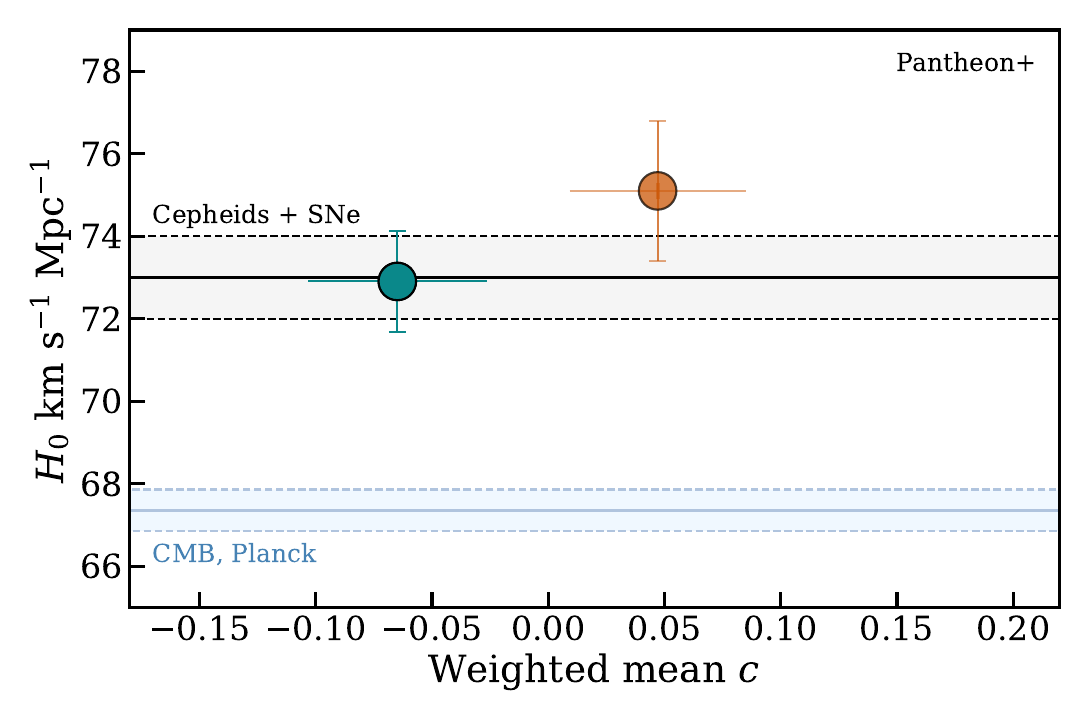}{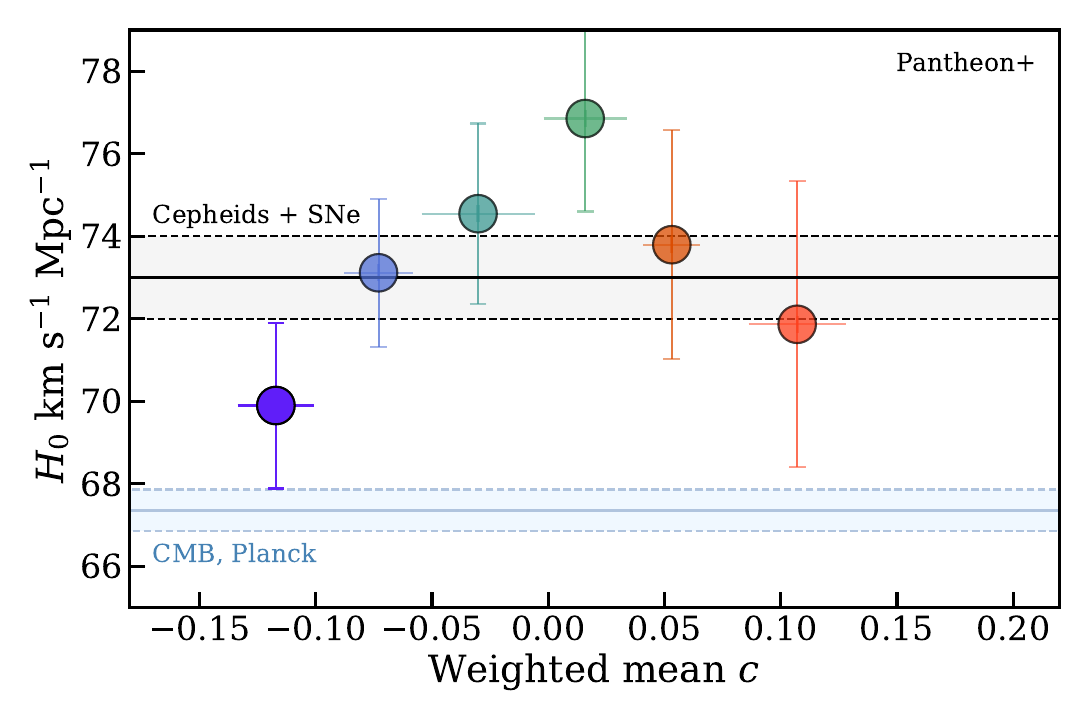}
\plottwo{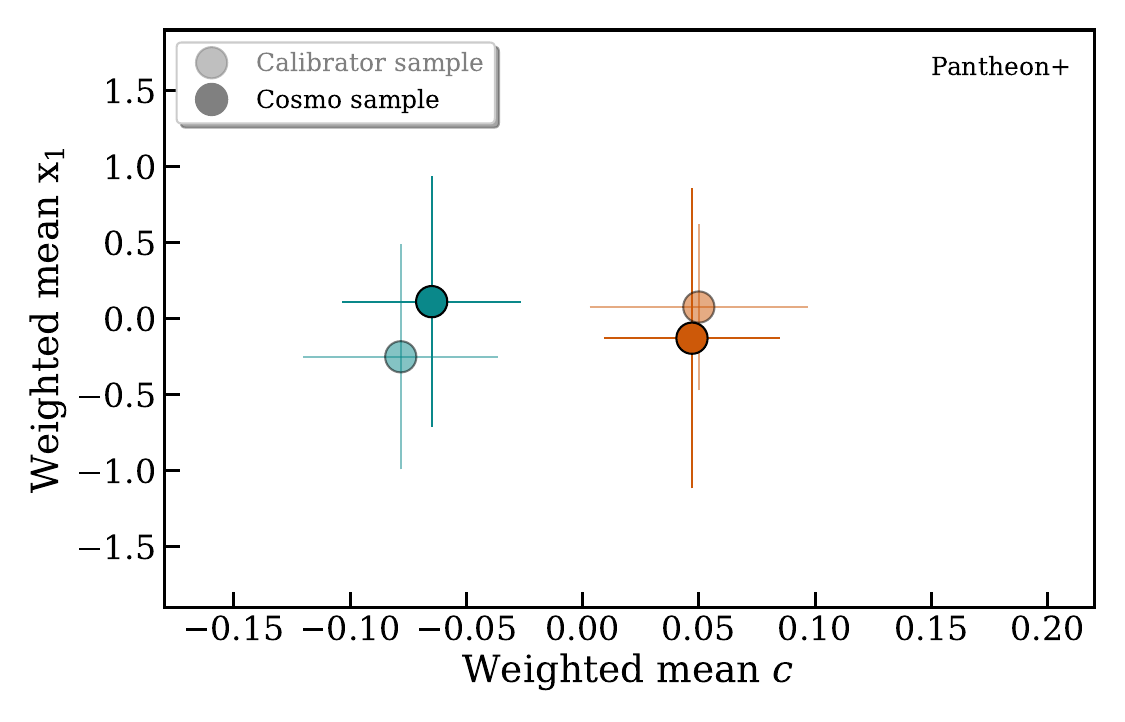}{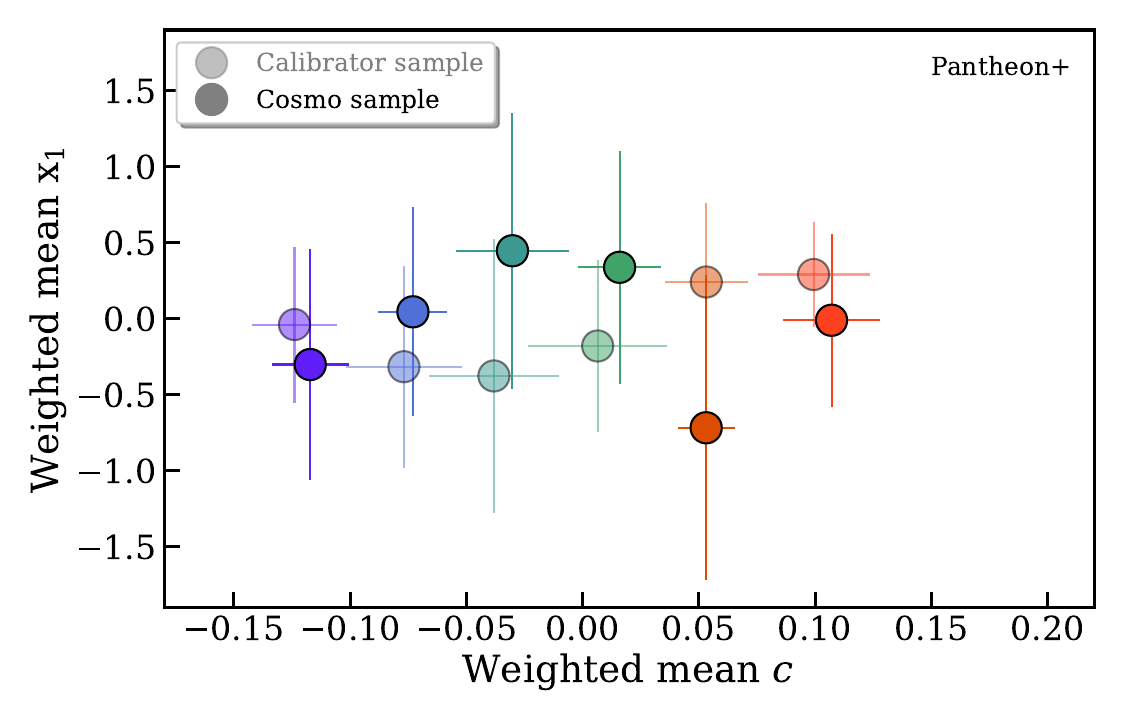}
\plottwo{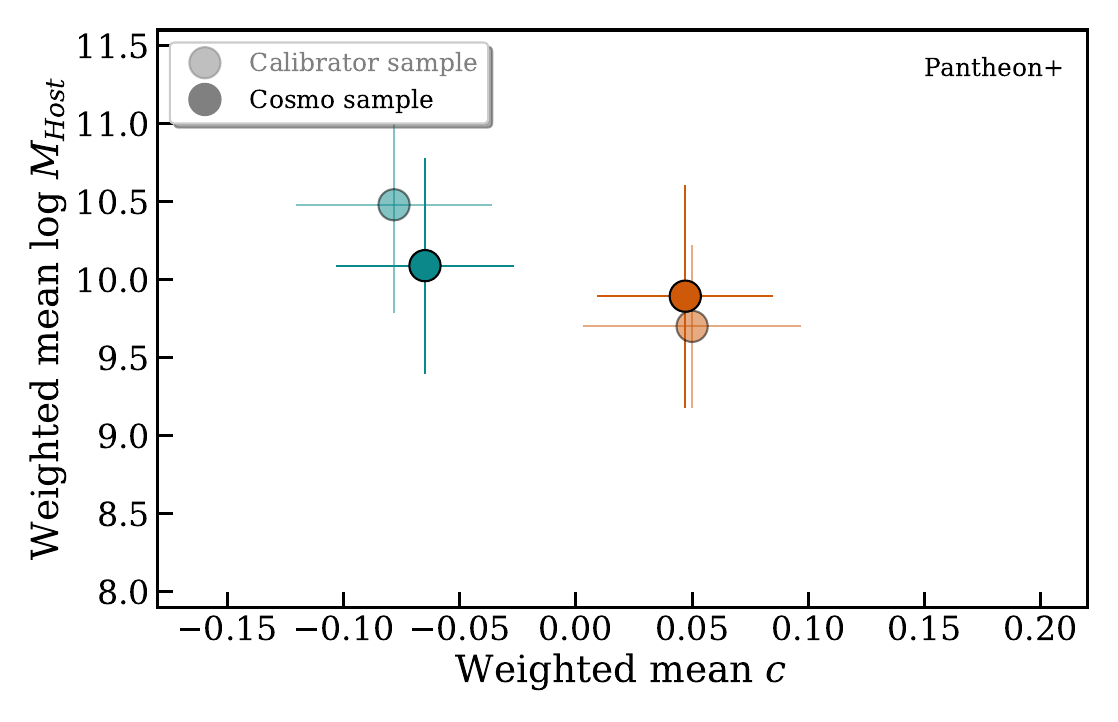}{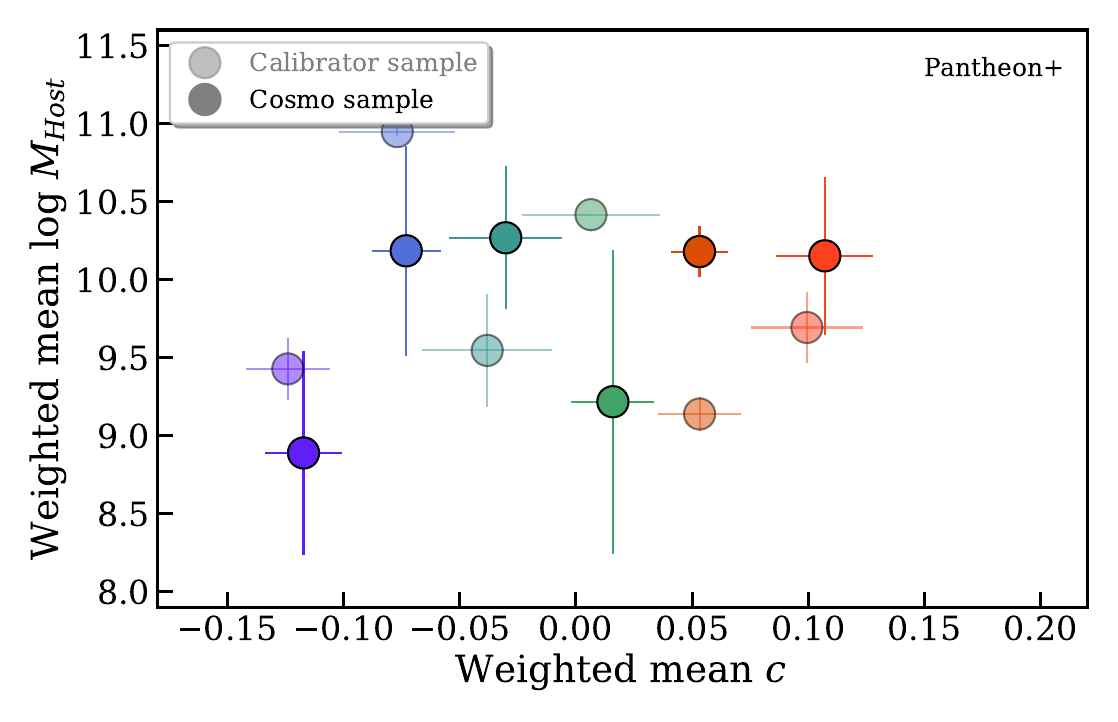}
\caption{$H_0$ and mean lightcurve parameters for different color bins of the Pantheon+ data compilation. The left column shows the results of the dataset split into colors of $c < 0$ and $c > 0$. The right column shows the results of dataset split into six equal color bins. {\it Upper panels:} show $H_0$. {\it Middle panels:} show the weighted mean SALT2 lightcurve shape parameter $x_1$ and standard deviation. {\it Lower panels:} show the weighted mean log host-galaxy mass and standard deviation. }
\label{fig:appcolbins}
\end{figure}

\begin{figure}[ht!]
\epsscale{1.0}
\plottwo{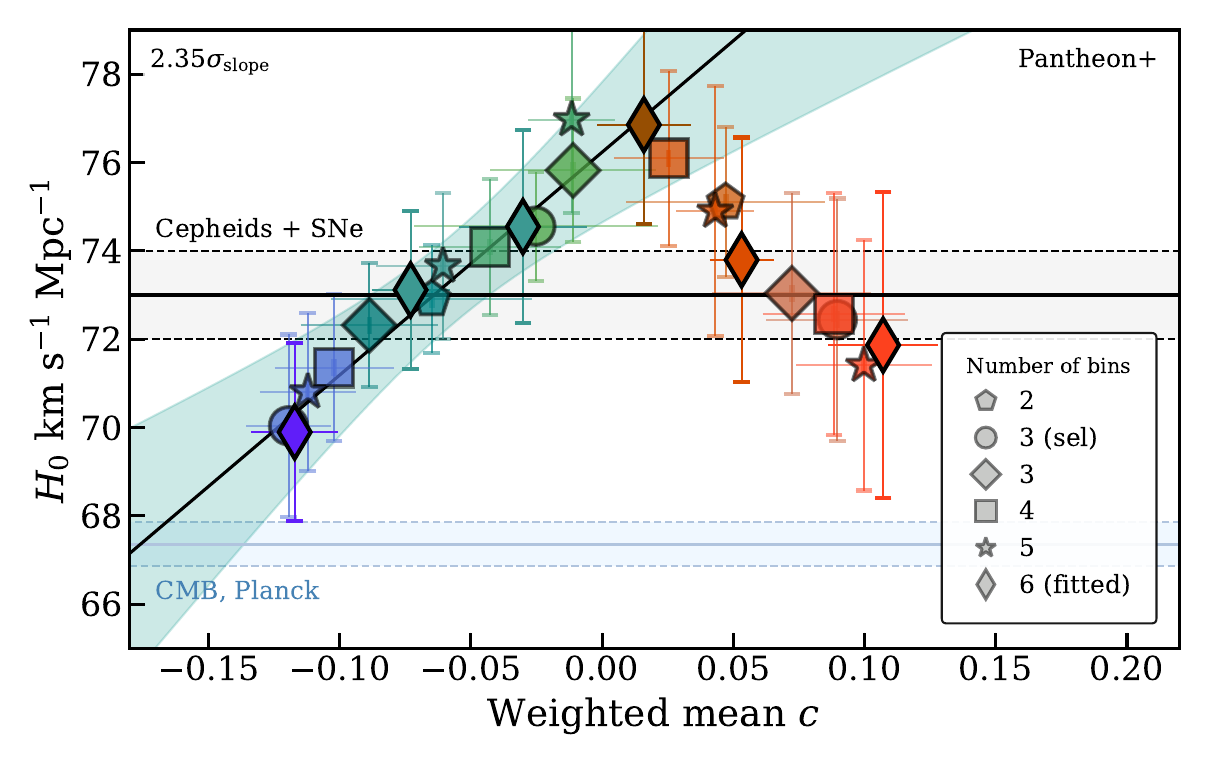}{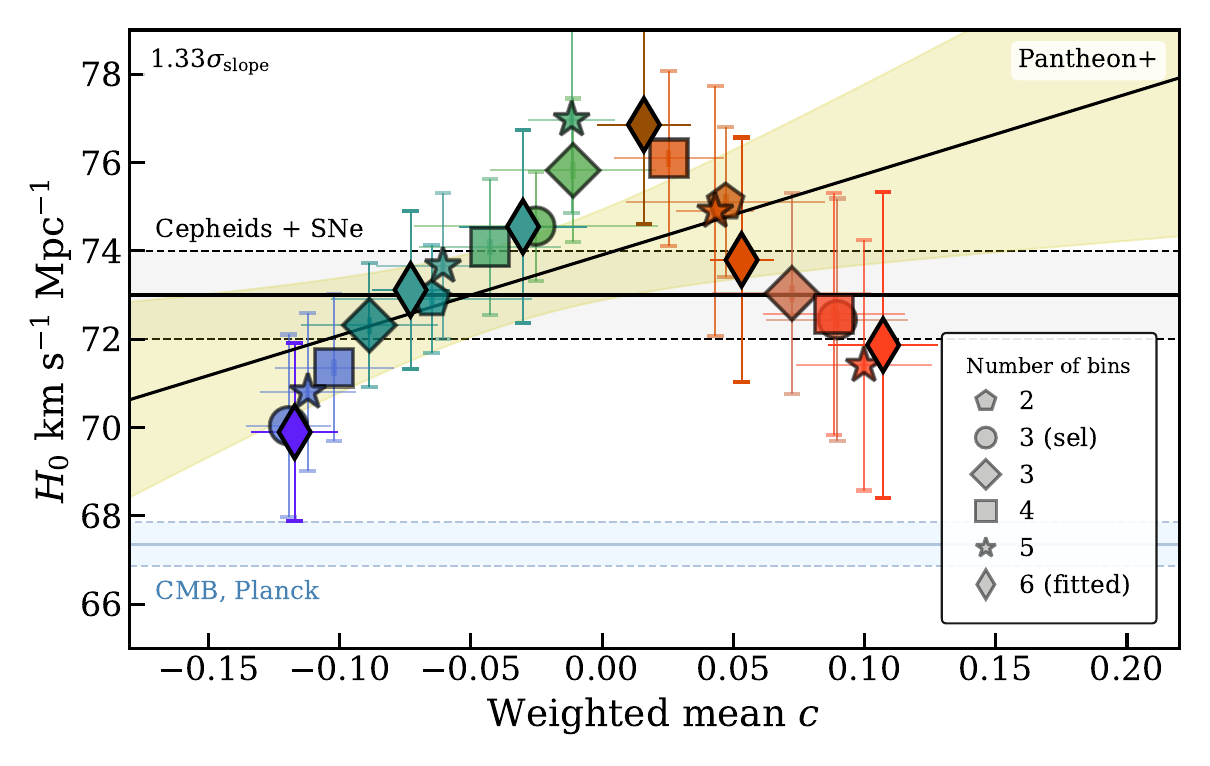}
\caption{Testing $H_0$ color dependence from the Pantheon+ data compilation. Shown are $H_0$ values for different selections of equal color bins, such as 2, 3, 4, 5, and 6 indicated as pentagon, diamond, square, star, and thin diamond symbols, respectively. For comparison, the three color bins as selected in Sect.~\ref{sec:bluetoredhubble} are shown as circles.    
The blue and yellow shaded regions and black solid lines through them are the weighted linear regressions of the inferred Hubble constant $H_0$ as a function of mean supernova color $c$ (6 bin case) using the pivot parameterization described in Section~\ref{Apsec:panth}.
{\it Left:} fit to the four bluest color bins, yielding a slope significance of $2.35\sigma_s$.
{\it Right:} fit to all six color bins, yielding a slope significance of $1.33\sigma_s$. 
}
\label{fig:appNbins_fit}
\end{figure}

Furthermore, to quantify the statistical significance of a potential dependence of $H_0$ on SN color, we perform a weighted linear regression of the form $H_0 = s c + p$, where $s$ denotes the slope and $p$ the intercept. The regression assumes independent $1\sigma$ uncertainties on $H_0$, which are incorporated through a weighted least-squares likelihood. Since each $H_0$ measurement is derived from a distinct color bin constructed to avoid overlap of individual SNe, the measurements are treated as statistically independent to first order.

Model parameters are estimated using the {\tt Python} routine {\tt curve\_fit} from the {\tt SciPy} optimization library \citep{2020NatMe..17..261V}, which determines best-fit parameters through minimization of the chi-square statistic,

\begin{equation}
\chi^2 = \sum_{i} \frac{\left(H_{0,i} - H_{0,\mathrm{model},i}\right)^2}{\sigma_{H_0,i}^2},
\end{equation}
where $H_{0,\mathrm{model},i} = s c_i + p$ and $\sigma_{H_0,i}$ is the $1\sigma$ uncertainty for each $H_{0,i}$.
The algorithm employs a nonlinear least-squares solver based on the Levenberg--Marquardt method \citep{1978LNM...630..105M}. Under the assumption of locally Gaussian likelihood contours, the parameter covariance matrix returned by the fit is used to derive parameter uncertainties and correlations.
To improve numerical stability and reduce covariance between the slope and intercept parameters, we adopt a pivoted form of the independent variable:

\begin{equation}
H_{0,{i}} = s \left(c_{i} - c_{0}\right) + p_c ,
\end{equation}
with $c_{0}$ as the mean color of the bins used in each regression, and $p_c$ is the intercept evaluated at the pivot color. The standard intercept corresponding to $c = 0$ is recovered as $p = p_c - s c_{0}$.

This approach allows for a numerically stable estimation of both $s$ and $p$, particularly when the dynamic range of $c$ is small compared to the scale of $H_0$. The slope $s$ represents the sensitivity of the inferred $H_0$ to the mean supernova color, providing a direct probe of potential color-dependent systematic effects rather than a physical causal relation.

The stability of the inferred slope was validated using Monte Carlo realizations in which $H_0$ measurements were perturbed according to their uncertainties. The dispersion of recovered slopes is consistent with the parameter uncertainty derived from the covariance matrix returned by the $\chi^2$ minimization,
supporting the validity of the Gaussian error approximation adopted in the regression.

Figure~\ref{fig:appNbins_fit} (left panel) shows the fit to the four bluest of six color bins, with a slope significance of 2.35$\sigma_s$, suggestive of a possible color dependence of $H_0$ on the blue end. Including all six bins  (Figure~\ref{fig:appNbins_fit}; right panel) results in a reduced slope significance of 1.33$\sigma_s$. A fit restricted to the three reddest bins results in a slope significance of 1.26$\sigma_s$, indicating that red SNe ($c>0$) are consistent with a constant high $H_0$ within uncertainties. All regression parameters are summarized in Table \ref{tab:pivot_regression_tests}.

\section{CSP sample, testing color dependence}
\label{Apsec:csp}

Similar to \csb\ in Sect.~\ref{Apsec:panth}, we show results for different color bins using the \csu\ sample.  Similar to above, we calculate $\Delta (B-V) = ((B-V) _{\rm cal, max} - (B-V) _{\rm cal, min}) / bin$,  which results in $\Delta (B-V)  \approx 0.16$ mag and $\Delta (B-V) \approx$  0.08 mag for bin=2 and bin=4, respectively. 
Furthermore, we test a simpler standardization in analogy to \citet{2021AA...647A..72K} with 
\begin{equation}
m_B = P^0 + P^1(s_{BV} - 1) + R(B_{\rm max} - V_{\rm max}) + \mu
\label{eq:apset1_mb}
\end{equation}
and
\begin{equation}
        \Gamma_{X} = \frac{(m_B - m_{B, model})^2}{\sigma_{X}^2}. 
\label{eq:apset1_gam}
\end{equation}
with $\sigma_{X}^2$, as defined in Eq. 10 and 11 in  \citet{2021AA...647A..72K}, which includes an additional term for the intrinsic scatter, $\sigma_{{\rm int},X}^2$. The total number of model parameters to be fit reduces to five ($P^0, P^1, R, \sigma_{\rm int}$ and $H_0$). 
We find that there is no significant difference in any of the fitted parameters of each color bin from either the complex model as employed in \citet{2024ApJ...970...72U} or the simpler standardization similar to what is employed in \citet{2021AA...647A..72K}.
A comparison of the results of $H_0$ and $R$ of the two models is shown in 
Figure~\ref{fig:appcolbinsUDDM24}, while the \csu\ model for four color bins is shown in Figure~\ref{fig:appcolbinsUDD}. Results of $H_0$ and $R$ for an alternative binning with blue and red color cuts obtained using a mean peak color difference between the color distributions of the HF samples (Figure~\ref{fig:distribution}) of 0.04 (see Sect.~\ref{sec:bluetoredhubble}), resulting in $c<-0.06$ (blue bin) and $c>0.1$ (red bin) with green in between (Figure~\ref{fig:appcolbinsUDDM22alt}).

\begin{figure}[ht!]
\epsscale{1.0}
\plottwo{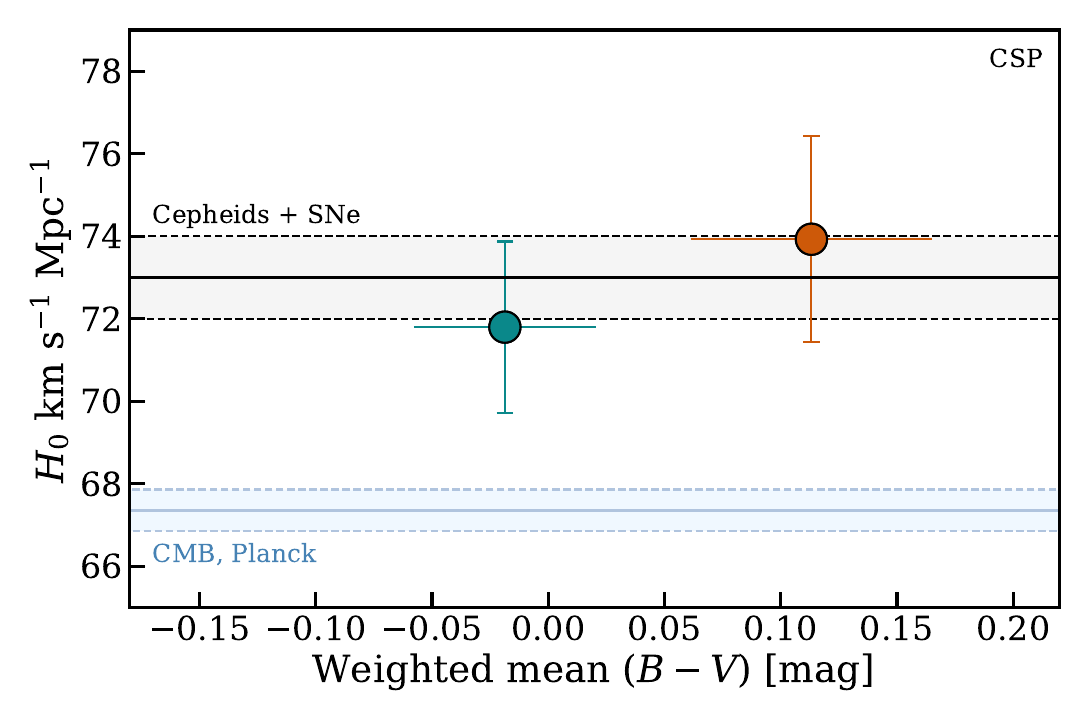}{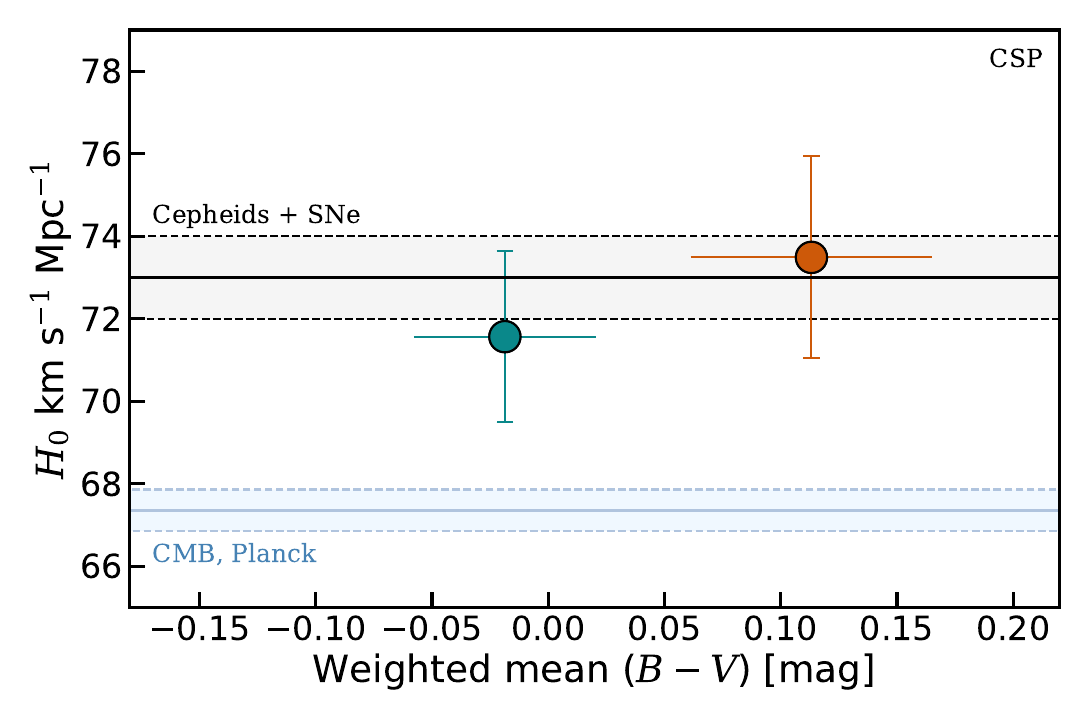}
\plottwo{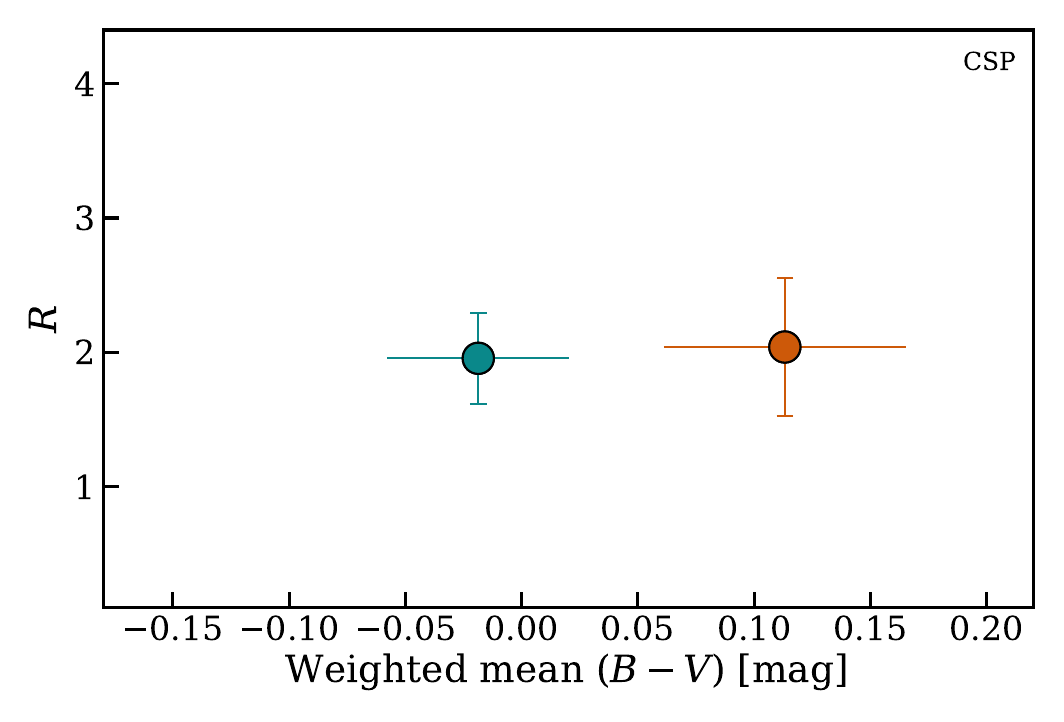}{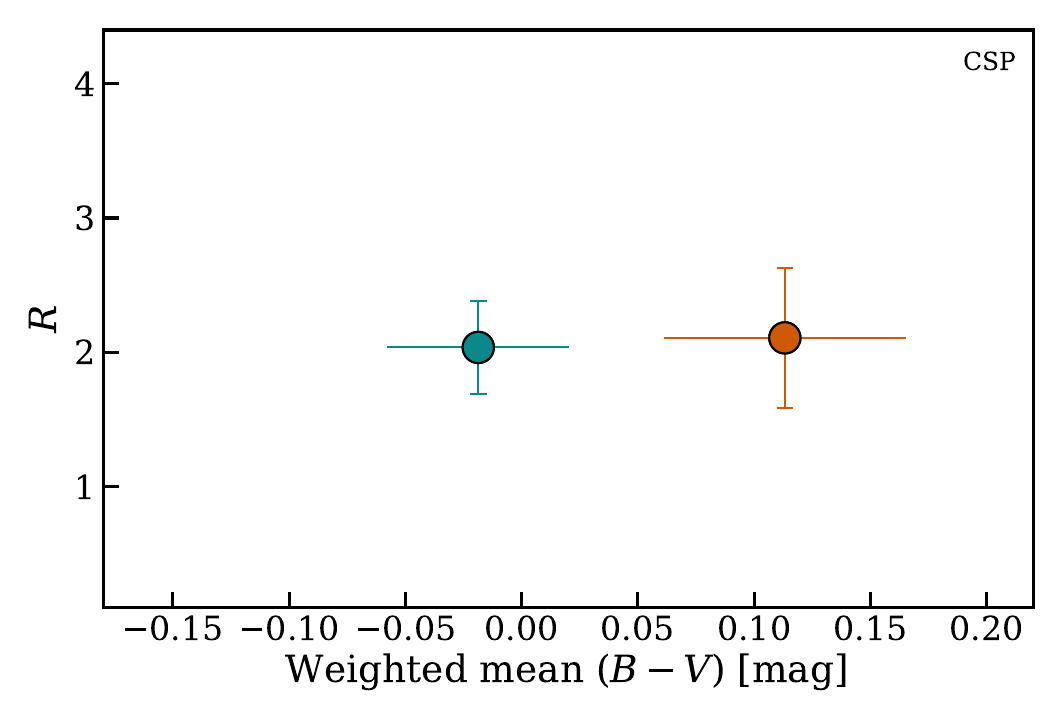}
\caption{$H_0$ and $R$ for two color bins and different models. The left column shows the results of the CSP data compilation split into color bins of $(B-V) < 0.0525$ and $(B-V) > 0.0525$. The right column shows the results of the same two color bins and for the simpler standardization as defined in Eq~\ref{eq:apset1_mb} and \ref{eq:apset1_gam}. }
\label{fig:appcolbinsUDDM24}
\end{figure}

\begin{figure}[ht!]
\epsscale{1.0}
\plottwo{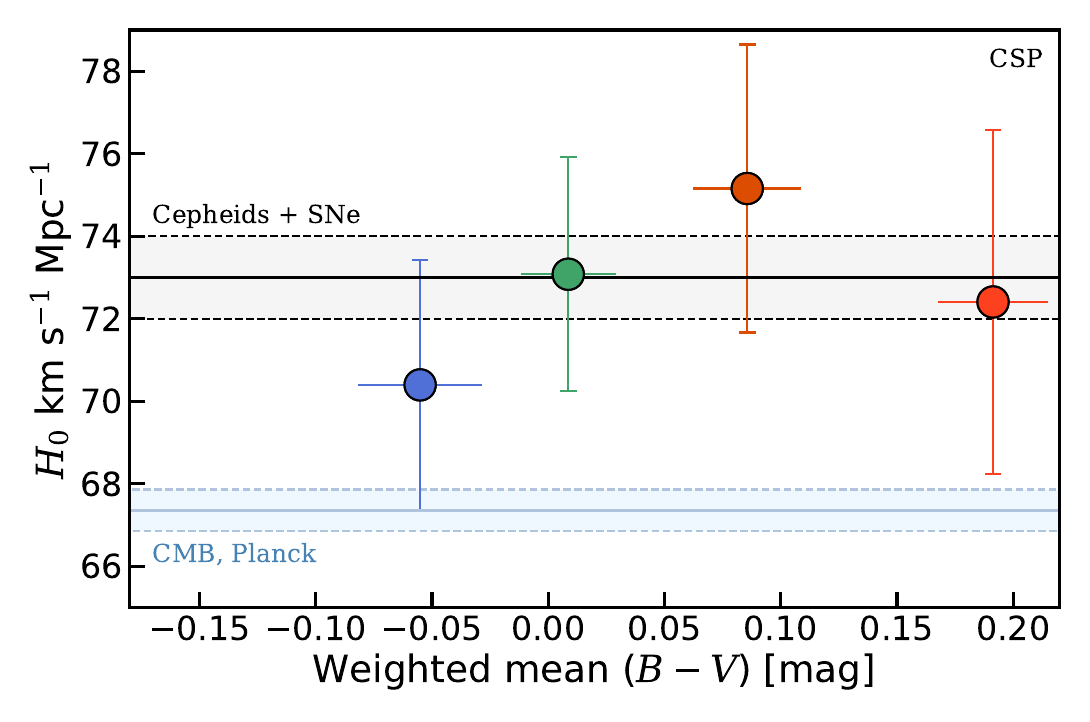}{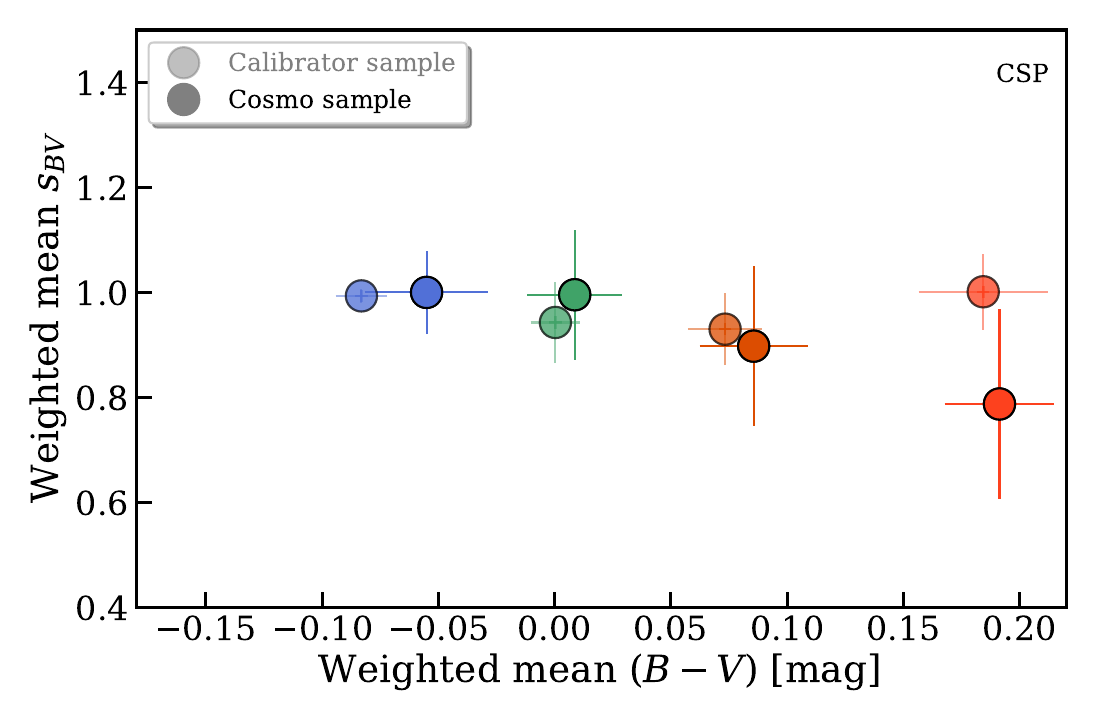}
\plottwo{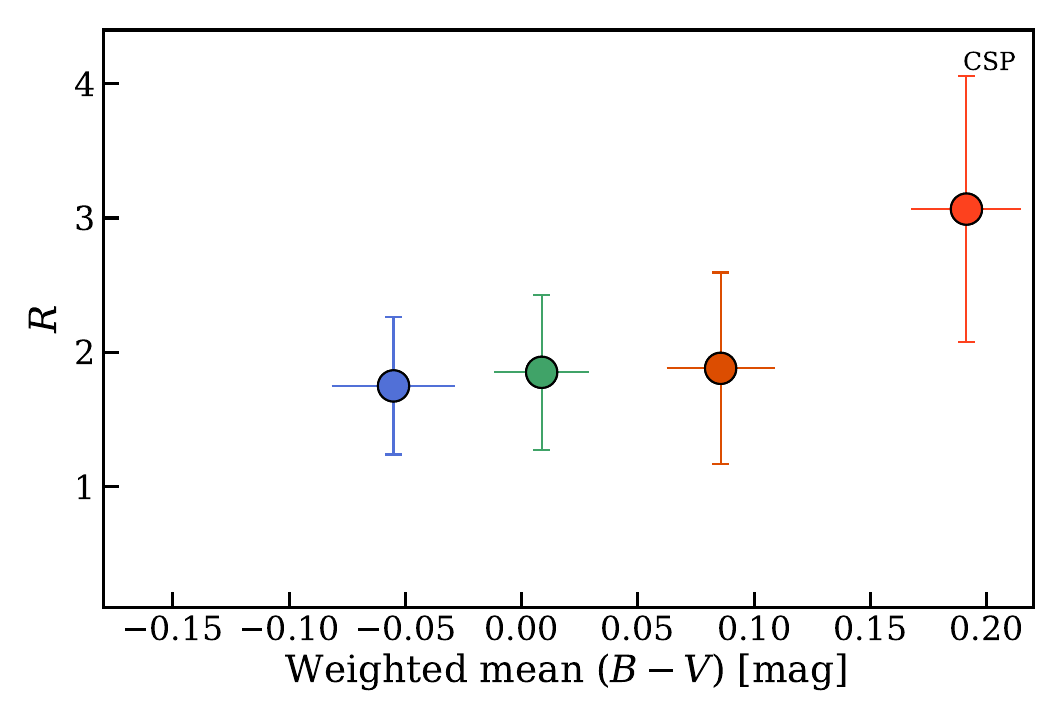}{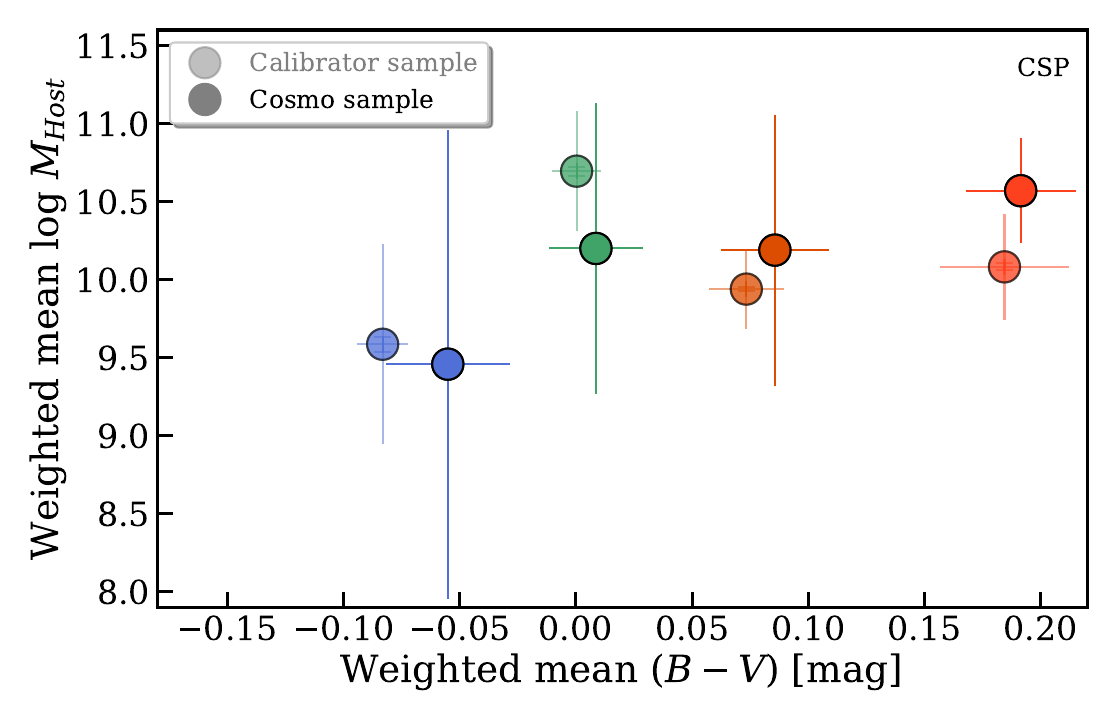}
\caption{$H_0$ and mean lightcurve parameters for four color bins of \csu. Shown are the results for $H_0$ (upper left) and $R$ (lower left), the weighted mean lightcurve shape parameter $s_{BV}$ and standard deviation (upper right), and the weighted mean log host-galaxy mass and standard deviation (lower right). }
\label{fig:appcolbinsUDD}
\end{figure}

\begin{figure}[ht!]
\epsscale{1.0}
\plottwo{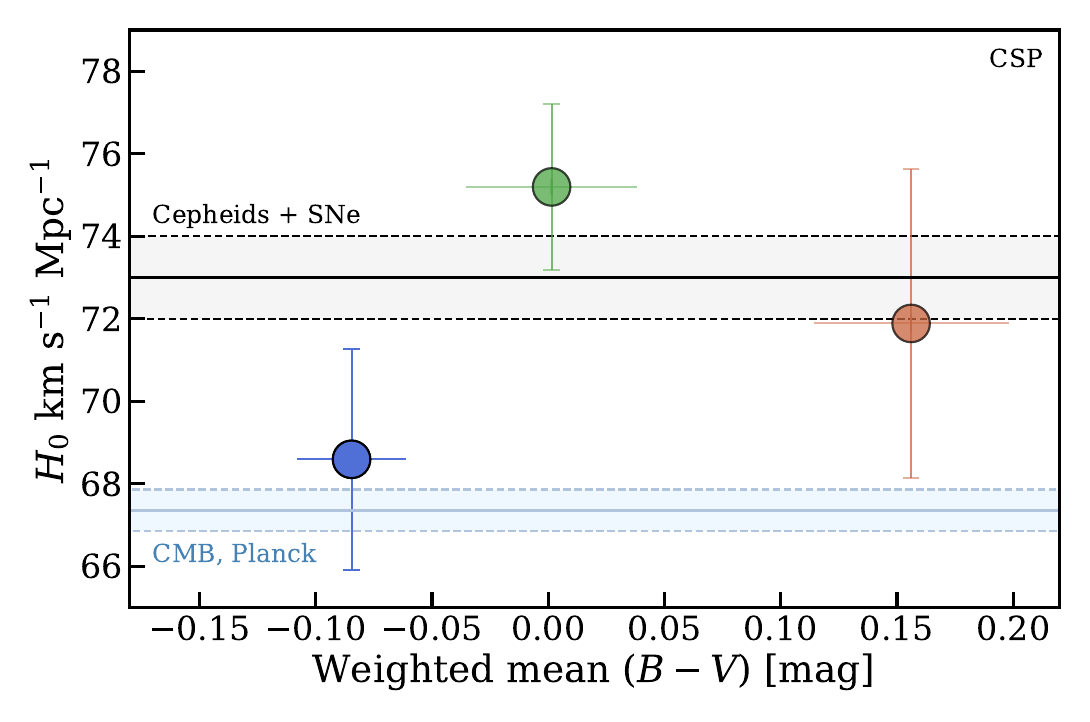}{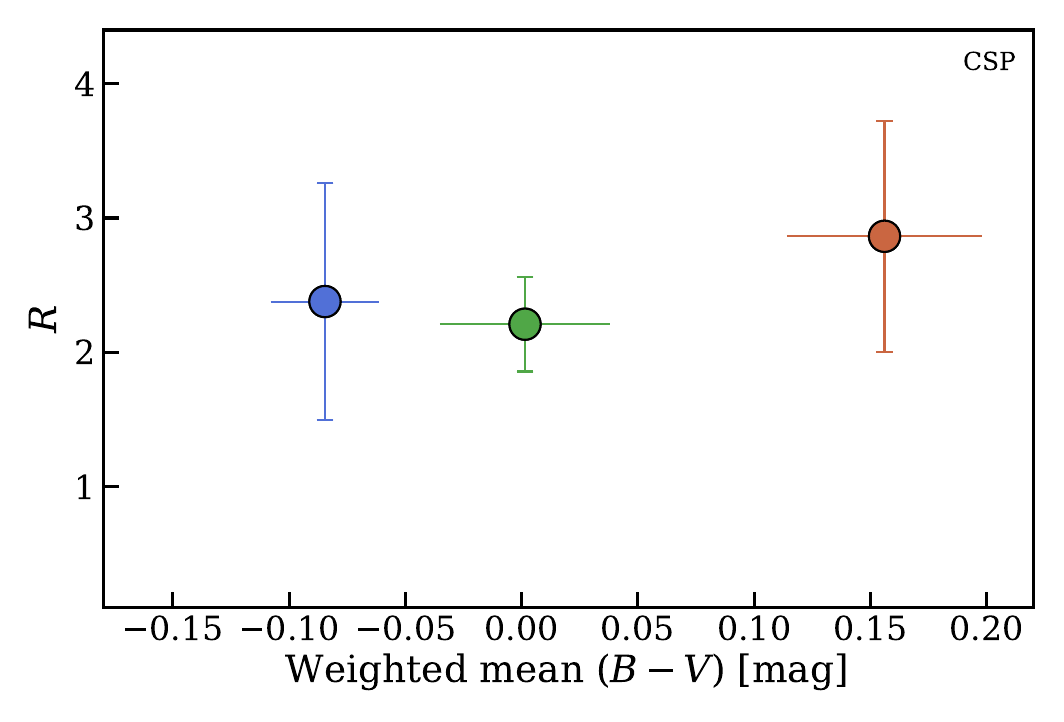}
\caption{Alternative binning for \csu. {\it Left panel:} shown are the results of $H_0$. {\it Right panel:} shown are the results of $R$. Color bin bounds are calculated using a mean peak color difference between the color distributions of the HF samples (Figure~\ref{fig:distribution}) of 0.04, resulting in $c<-0.06$ (blue bin) and $c>0.1$ (red bin) with green in between.}
\label{fig:appcolbinsUDDM22alt}
\end{figure}

\subsection{Testing Eddington Bias}
As discussed in Sec.~\ref{sec:bluetoredhubble}, here we test for Eddington bias using the same \csu\ setup with which results have been obtained, as shown in Figure \ref{fig:hnull1}. However, we only select SNe~Ia which are including uncertainties within a color bin. This obviously leads to a reduction of the number of SNe~Ia in each bin. 
The results, in comparison to what is also shown in Figure~\ref{fig:hnull1} are shown in Figure~\ref{fig:appcolbinsUDDM2edd}.

\begin{figure}[ht!]
\epsscale{1.0}
\plottwo{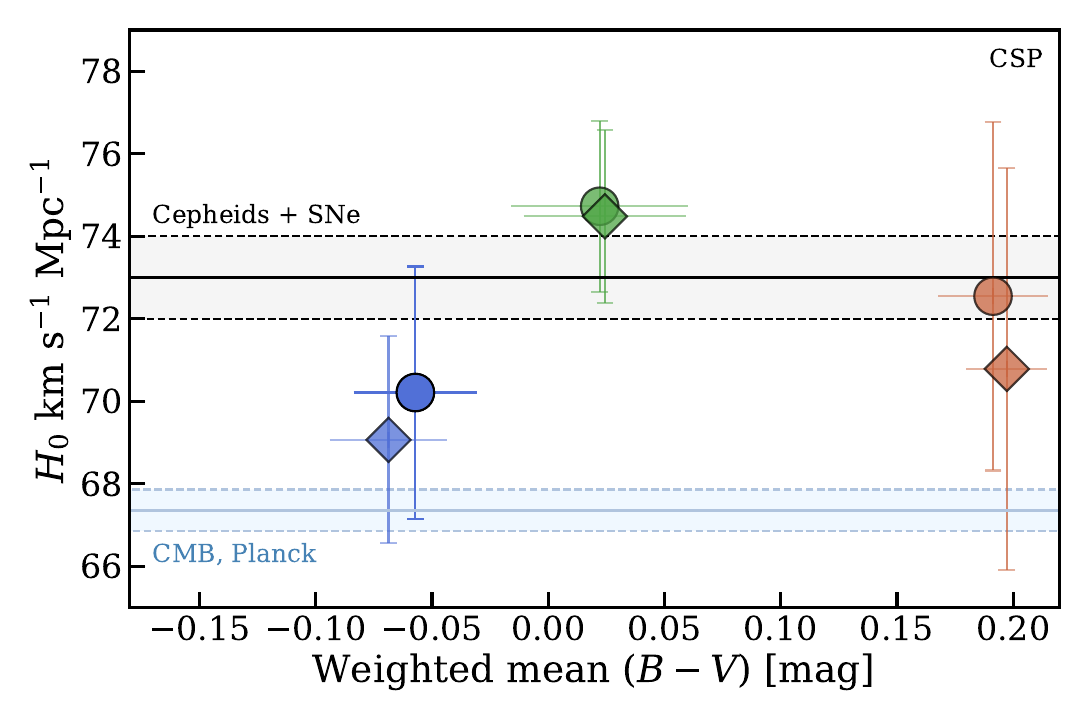}{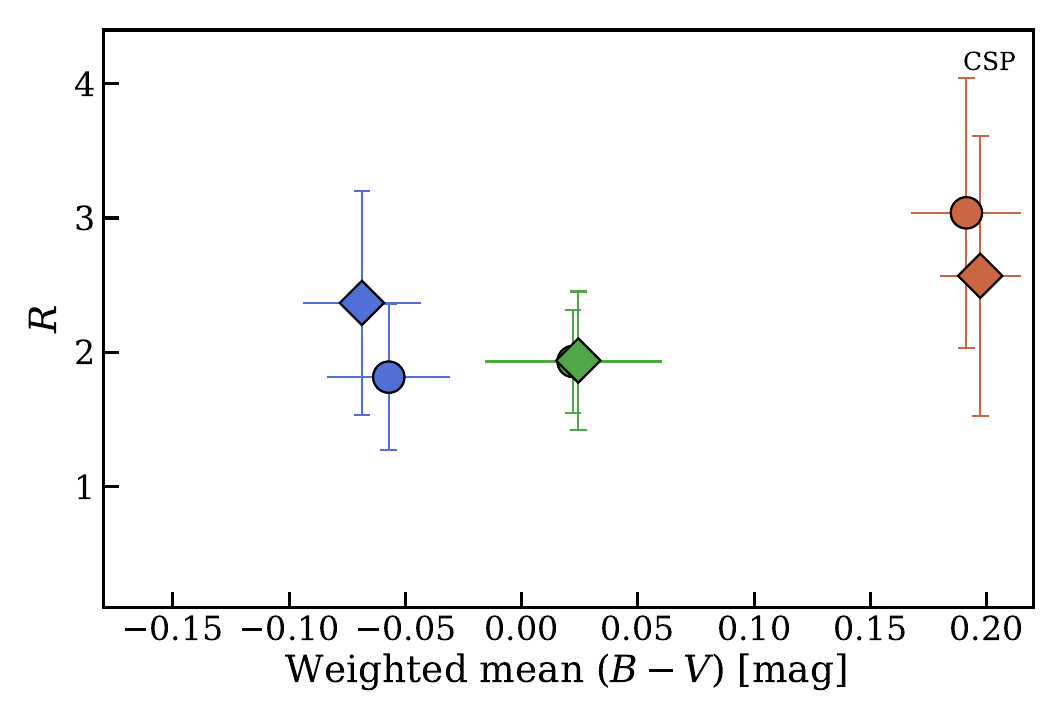}
\caption{Testing Eddington bias for \csu. Shown are the results $H_0$ (left panel) and $R$ (right panel). Squared symbols mark the model with SNe~Ia within a color bin, including their uncertainties.}
\label{fig:appcolbinsUDDM2edd}
\end{figure}

\section{Tables}

\begin{deluxetable*}{lccccc}[htb!]
\label{tab:PANTHdata}
\tablecaption{
 {\bf Cosmological Tests with Different Data Sets: \csb} }
\linespread{1.1}\selectfont\centering
\tablewidth{0pt}
\tablehead{
\colhead{Bin} &
\colhead{$M_b$}     &
\colhead{$H_0$} 
} 
\startdata 
\noalign{\smallskip}
\hline
\noalign{\smallskip}
Blue & $-$19.32(0.06) & 69.98(2.08) \\ 
Green & $-$19.23(0.03) & 74.51(1.23)\\
Red & $-$19.27(0.07) & 72.47(2.72)  \\ 
\hline
\enddata
\end{deluxetable*}
\begin{deluxetable*}{lccccccc}[htb!]
\label{tab:UDDdata}
\tablecaption{
 {\bf Cosmological Tests with Different Data Sets: \csu} }
\linespread{1.1}\selectfont\centering
\tablewidth{0pt}
\tablehead{
\colhead{Bin} &
\colhead{$P^0$}     &
\colhead{$P^1$}     &
\colhead{$P^2$}     &
\colhead{$H_0$} &
\colhead{$R$}    &
\colhead{$\sigma_{\rm int}$}   &
\colhead{$vel_{\rm pec}$}
} 
\startdata 
\noalign{\smallskip}
\hline
\noalign{\smallskip}
Blue  & $-$19.24(0.10)  & $-$1.26(0.26) & 2.0(1.77) & 70.27(3.0) & 1.8(0.55) & 0.16(0.03) & 525(319)\\ 
Green & $-$19.06(0.06) & $-$1.00(0.14) & 0.65(0.80) & 74.73(2.1) & 1.9(0.40) & 0.19(0.01) & 180(133)\\
Red   & $-$19.25(0.19) & $-$0.85(0.42) & 0.39(2.02) & 72.55(4.2) & 3.04(1.00) & 0.18(0.05) & 713(493)\\ 
\hline
\enddata
\end{deluxetable*}

\begin{deluxetable*}{lcccccccccc}
\tablecaption{\bf Results of Pivot Linear Regression Tests for the Dependence of $H_0$ on SN Color. \label{tab:pivot_regression_tests}}
\tablehead{
\colhead{Test} &
\colhead{$\#$ Bins} &
\colhead{Slope $s$} &
\colhead{$\sigma_s$} &
\colhead{$c_{\rm piv}$} &
\colhead{$H_{0,\rm piv}$} &
\colhead{$\sigma_{H_{0,\rm piv}}$} &
\colhead{$p$ ($c$=0)} &
\colhead{$\sigma_p$} &
\colhead{Significance} &
\colhead{Significance (MC)}
}
\startdata
Blue bins &  4 & $50.46$ & $21.46$ & $-0.0569$ & $73.36$ & $1.02$ & $76.23$ & $1.59$ & $2.35$ & $2.34$ \\
Red bins & 3 & $-56.58$ & $44.93$ & $0.0462$ & $74.87$ & $1.56$ & $77.49$ & $2.60$ & $1.26$ & $1.26$ \\
All bins & 6 & $18.22$ & $13.71$ & $-0.0331$ & $73.31$ & $0.92$ & $73.91$ & $1.03$ & $1.33$ & $1.32$ \\
\enddata
\tablecomments{
$c_{\rm piv}$ denotes the pivot color corresponding to the weighted mean color of the SN sample used in each test.
$H_{0,\rm piv}$ represents the inferred Hubble constant evaluated at $c_{\rm piv}$.
The parameter $p$ corresponds to the standard intercept evaluated at $c=0$.
Analytic uncertainties are derived from the covariance matrix of the weighted least-squares fit.
Monte Carlo (MC) uncertainties are obtained from realizations of $H_0$ measurements perturbed according to their reported uncertainties.
Significances are defined as $s / \sigma_s$.
}
\end{deluxetable*}

\end{document}